\pdfoutput=1
\documentclass[preprintnumbers,amsmath,amssymb,prd,notitlepage,nofootinbib,twocolumn,superscriptaddress,table,xcdraw,floatfix]{revtex4-2}

\usepackage{graphicx}
\usepackage{dcolumn}
\usepackage{bm}
\usepackage{mathtools}
\usepackage{wasysym}
\usepackage{adjustbox}
\usepackage{verbatim}
\usepackage[dvipsnames]{xcolor}
\usepackage{hyperref}
\usepackage{todonotes}
\usepackage{float}

\usepackage{aasmacros}
\usepackage{xspace}
\usepackage{tabularx}
\usepackage[font=footnotesize, caption=false]{subfig}
\usepackage{babel}
\usepackage{blindtext}
\usepackage{verbatim}

\usepackage{colortbl}
\mathchardef\mhyphen="2D

\makeatletter

    \def\CT@@do@color{%
      \global\let\CT@do@color\relax
            \@tempdima\wd\z@
            \advance\@tempdima\@tempdimb
            \advance\@tempdima\@tempdimc
    \advance\@tempdimb\tabcolsep
    \advance\@tempdimc\tabcolsep
    \advance\@tempdima2\tabcolsep
            \kern-\@tempdimb
            \leaders\vrule
                    \hskip\@tempdima\@plus  1fill
            \kern-\@tempdimc
            \hskip-\wd\z@ \@plus -1fill }
\makeatother

\def\trans{\mathsf{T}}

\newcommand{\Neff}{N_\mathrm{eff}}

\newcommand{\smu}{Department of Physics,
Southern Methodist University, 
Dallas, TX 75275, USA}

\begin{document}

\title{Estimating the Impact of foregrounds on the Future Detection of Rayleigh scattering}

\author{Yijie Zhu}
\affiliation{Department of Physics and Astronomy, Stony Brook University, Stony Brook, NY 11794, USA}
\affiliation{Department of Physics, Cornell University, Ithaca, NY 14853, USA}
\author{Benjamin Beringue}
\affiliation{School of Physics and Astronomy, Cardiff University, The Parade, Cardiff, Wales CF24 3AA, UK}
\affiliation{DAMTP, Centre for Mathematical Sciences, Wilberforce Road, Cambridge, UK, CB3 0WA}
\author{Steve K. Choi}
\affiliation{Department of Physics, Cornell University, Ithaca, NY 14853, USA}
\affiliation{Department of Astronomy, Cornell University, Ithaca, NY 14853, USA}
\author{Nicholas Battaglia}
\affiliation{Department of Physics, Cornell University, Ithaca, NY 14853, USA}
\author{P. Daniel Meerburg}
\affiliation{Van Swinderen Institute for Particle Physics and Gravity, University of Groningen, Nijenborgh 4, 9747 AG Groningen, The Netherlands}
\author{Joel Meyers}
\affiliation{\smu}

\begin{abstract}
Rayleigh scattering of the cosmic microwave background (CMB) by neutral hydrogen shortly after recombination leaves frequency-dependent imprints on intensity and polarization fluctuations. High signal-to-noise observations of CMB Rayleigh scattering would provide additional insight into the physics of recombination, including greater constraining power for parameters like the primordial helium fraction, the light relic density, and the sum of neutrino masses. However, such a measurement of CMB Rayleigh scattering is challenging due to the presence of astrophysical foregrounds, which are more intense at the high frequencies, where the effects of Rayleigh scattering are most prominent. Here we forecast the detectability of CMB Rayleigh scattering including foreground removal using blind internal linear combination methods for a set of near-future surveys. We show that atmospheric effects for ground-based observatories and astrophysical foregrounds pose a significant hindrance to detecting CMB Rayleigh scattering with experiments planned for this decade, though a high-significance measurement should be possible with a future CMB satellite.

\end{abstract}

\maketitle

\section{Introduction}

The early universe was filled with a hot dense plasma composed primarily of electrons, protons, and helium nuclei in a bath of radiation tightly coupled to the plasma due to the high rate of Thomson scattering. As the universe cooled due to cosmic expansion, the plasma recombined to form a transparent gas of neutral hydrogen and helium. Shortly after recombination, the photons decoupled and have traveled mostly unimpeded ever since. These photons, which continued to redshift due to the cosmic expansion, make up the cosmic microwave background (CMB).

Observations of the CMB have provided a wealth of information about the history, contents, and evolution of the universe. The existence of a thermal bath of radiation provides very strong evidence for the Hot Big Bang model of cosmology, and measurements of the CMB temperature and polarization anisotropies have allowed for percent-level determination of the parameters that define the current $\Lambda$CDM model of cosmology~\cite{Planck2018cosmo}. Future CMB surveys will extract a great deal more information from the temperature and polarization anisotropies.

Looking beyond primary CMB anisotropies, additional processes after recombination leave imprints on the CMB. One such process is Rayleigh scattering of CMB photons by neutral atoms, especially by hydrogen, which achieves maximum density shortly after recombination. Rayleigh scattering is a classical scattering process of long wavelength radiation by the induced dipole of neutral species. Its impact on recombination and CMB anisotropies has been studied in \citet{Lewis2013,Yu2001, Alipour2014}. The spectral function for Rayleigh is well described by a cross-section that scales as $\nu^4$. However, as we will show, observing Rayleigh scattering of the CMB remains challenging due to the presence of astrophysical foregrounds that act as a source of confusion at the relevant frequencies.

Physically, Rayleigh scattering causes a frequency-dependent damping of CMB anisotropies by scattering photons from hot spots out of the line-of-sight and from cold spots into the line of sight. At high frequencies the effect of Rayleigh scattering contributes to the angular power spectrum, and thus it must be consistently included in cosmological modeling. 

Observing the effects of Rayleigh scattering of the CMB would provide additional information that can be used to probe the early universe, including the physics of recombination,  the ionization history, the expansion rate of the early universe, and the properties of primordial perturbations. One interesting application of CMB Rayleigh scattering is to improve measurements of the light relic density ($\Neff$), which further probes physics beyond the Standard Model in the dark sector \citep{Green2019,Dvorkin:2022jyg}. With primary CMB observations alone, measuring $\Neff$ requires significantly more resources when compared to including Rayleigh scattering information \citep[e.g.,][]{CMBS4,Beringue2020}. However, the value of Rayleigh in constraining physics beyond the Standard Model requires a significant detection of the signal. For example, for experiments such as CCAT-prime~\citep{CCATp}, the Simons Observatory (SO)~\citep{SOforecasts}, and the \textit{Planck} satellite~\citep{Planck2018}, even under ideal circumstance, where foregrounds and other possible systematic effects are neglected, the expected statistical significance of the Rayleigh scattering signal would rise to the level of a detection, but would not significantly improve constraints on cosmological parameters~\citep{Beringue2020}.

The amplitude of the Rayleigh auto-spectrum is very small, both in temperature and in polarization. At around $\nu=500~\mathrm{GHz}$, the temperature Rayleigh scattering auto-spectrum is still around $7$ orders of magnitude lower than the primary CMB spectrum~\citep[][]{Lewis2013}. Hence a detection of the auto-spectrum will remain out of reach for the foreseeable future. On the other hand, cross-correlations between the Rayleigh scattering signal and the primary temperature or polarization anisotropies are more promising~\citep{Lewis2013,Alipour2014,Beringue2020}. The amplitude of the cross signals is around $3$ orders of magnitude larger than the auto-spectrum of Rayleigh scattering, as shown in Fig.~\ref{contribution}. Raw noise levels from current ground-based CMB experiments, the \textit{Planck} satellite and future CMB satellite missions~\citep[e.g., PICO][]{PICO} suggests the Rayleigh scattering cross-correlations should be detectable~\citep[e.g.,][]{Lewis2013,Alipour2014,Beringue2020}. The main challenge is to isolate the Rayleigh scattering signal in the presence of foregrounds and the atmosphere, both of which can have a large impact on detectability. While the role of the atmosphere is only relevant for ground-based experiments, astrophysical foregrounds will affect both satellite missions and ground-based experiments. This paper aims to address the effect of foregrounds on detectability of the Rayleigh scattering signal. Indeed, the frequency dependence of some foreground emissions make their amplitude significant in the frequency range where the Rayleigh scattering is brighter. It is timely to assess how large this impact will be for future CMB experiments.

Previously, \citet{Alipour2014} studied the effect of foregrounds and foregrounds removal on the Rayleigh scattering signal and concluded that the $TE$ cross-power spectra between primary temperature and Rayleigh scattering polarization signals should be relatively unbiased by foregrounds. More recently, \citet{Beringue2020} forecasted a detection of the $TT$ cross-correlation with $4.7\sigma$ for \textit{Planck}, and a whopping $715\sigma$ detection with PICO, both in the absence of foregrounds~\citep{Lewis2013,Beringue2020}. Throughout this paper, we denote the signals as $XY=\mathrm{CMB}\;(T/E)\times \mathrm{RS}\;(T/E)$, where $T$ represents the temperature fluctuations, $E$ refers to the $E$-mode polarization, and RS denotes the Rayleigh scattering component.
These calculations show that $TT$ cross-correlation signal-to-noise ratio is roughly an order of magnitude larger than the $TE$ or $ET$  signal-to-noise. Therefore, given the large difference in signal-to-noise in the absence of foregrounds, a more comprehensive study of all Rayleigh scattering cross-correlations, including foreground removal is warranted.

We use a component separation method to quantify the impact of foreground removal on the Rayleigh scattering signal-to-noise. Several component separation methods and algorithms already exist in the literature, including {\it blind} methods like Spectal Matching Independent Component Analysis (SMICA)~\citep[][]{2003MNRAS.346.1089D}, which was used in \textit{Planck} data analysis \citet{2014A&A...571A..12P}, Fast Independent Component Analysis (FastICA)~\citep[][]{2002MNRAS.334...53M}, the Joint Approximate Diagonalization of Eigenmatrices
method (JADE)~\citep[][]{2006A&A...455..741P}, and Correlated Component Analysis (CCA)~\citep[][]{2005EJASP2005.2400B}.
We chose to be conservative in our approach to component separation and selected the method known as Internal Linear Combination (ILC)~\citep[][]{Delabrouille2009}, which does not assume particular parametrization of the foreground components or templates.

Since we are dealing with the cross-spectra of two signals, we use a variation of this method, the constrained Internal Linear Combination (cILC)~\citep{Remazeilles_2010}. With this method, we avoid residual bias from the CMB primary on the Rayleigh signal, and vice versa in the cross-spectra \citep[][]{RemChluba2018}. Given the conservative nature of our approach, these forecasts should be considered as lower limits on the possible signal-to-noise.

In this paper, we introduce the physical modeling and scientific significance of the Rayleigh scattering signal in Section~\ref{sec:RSPhys}. We set up our modeling of the foregrounds and describe the experimental configurations we consider in Section~\ref{sec:fgmodel} and Section~\ref{sec:Tele}; additional information about the surveys can be found in Appendix~\ref{sec:Telescopes}. The CMB and Rayleigh spectra are generated used a modified version of \texttt{CAMB}\footnote{\url{https://camb.info/}}~\citep{Lewis:1999bs}. Details on the cILC method are presented in Section ~\ref{sec:ILC}. In Section~\ref{sec:Results} we forecast the signal-to-noise for CCAT-prime, SO and \textit{Planck}. In addition, we explore the effect of systematics: bias and gain calibration. We show how foregrounds impact our observation. Then, we perform the same analysis for PICO and explore combinations of data from LiteBIRD with CCAT-prime and \textit{Planck}. We use the extrapolated atmosphere noise model on all frequencies from ACT~\citep[][]{D_nner_2012}. The atmospheric noise for CCAT-prime and SO might differ from these estimates, so our results should be considered with this in mind. We sometimes use $\mathcal{D}_{\ell}$ which is related to ${C}_{\ell}$ as:
\begin{equation}
  \mathcal{D}_{\ell}=\frac{\ell(\ell+1)}{2\pi}C_{\ell}\, .
\end{equation}

\section{Rayleigh scattering Physics}
\label{sec:RSPhys}
Rayleigh scattering describes the classical scattering of electromagnetic waves by particles with size is much smaller than the wavelength of the radiation. To first order, the cross section of this scattering process is proportional to $\nu^4$, where $\nu$ is the photon frequency. Its impact on the recombination history and decoupling of CMB photons has been described in details in Refs.~\citep{Yu2001,Alipour2014,Lewis2013}. We will summarise the cosmological implications of Rayleigh scattering in this section. 

In the early universe, photons are tightly coupled with the free electrons in the primordial plasma by Thomson scattering. As the universe expands, the temperature decreases until recombination of free electrons and protons becomes thermally favored. As the fraction of free electrons drops, the Thomson scattering rate falls. Around redshift $z\sim 1100$, photons experience a last scattering event and start free-streaming. 

\begin{figure}[t!]

  \centering

  \includegraphics[width=\linewidth]{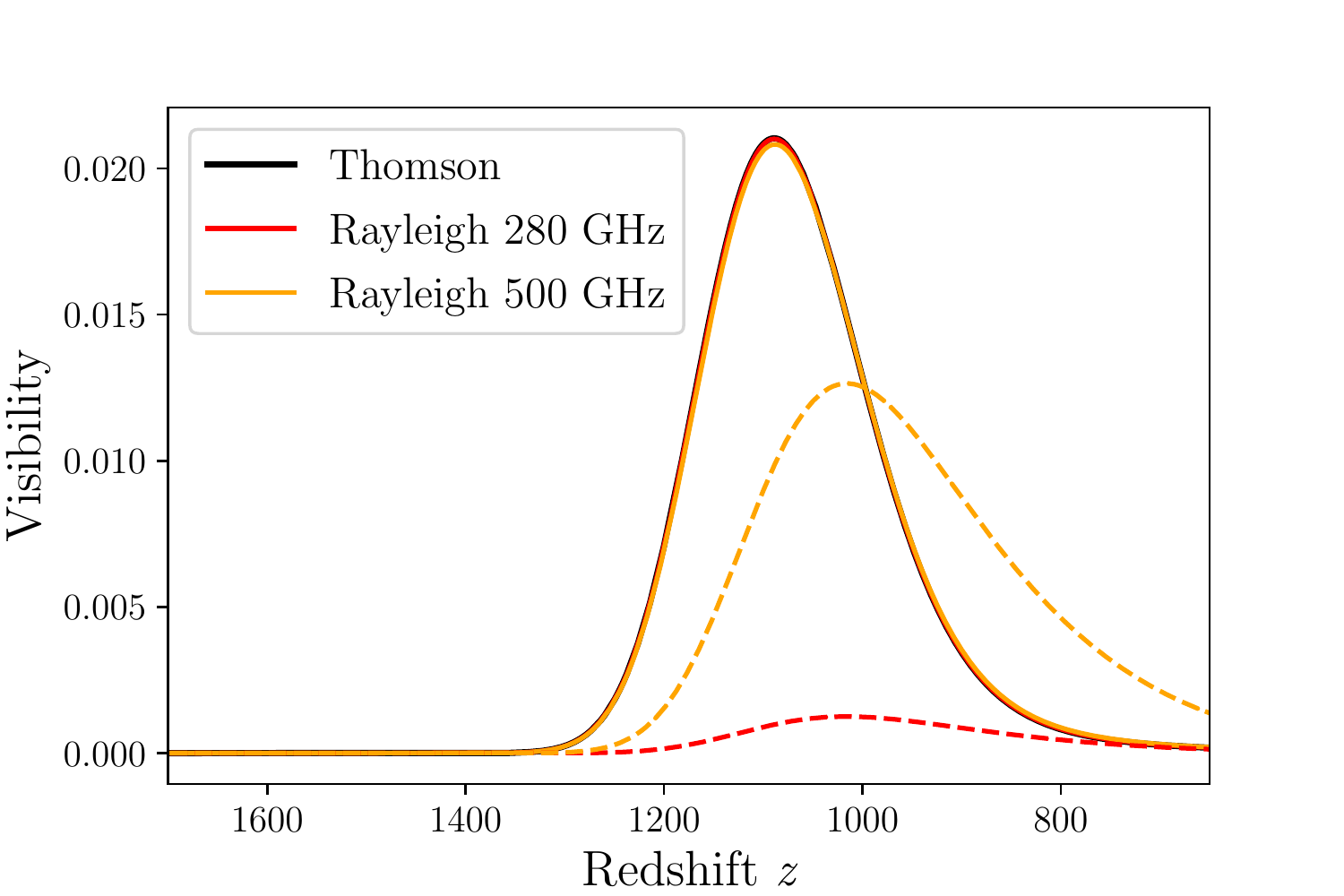}
 
\caption{Change to the visibility function $g(\tau)=\Dot{\tau}e^{-\tau}$ due to Rayleigh scattering at $\nu=280~\mathrm{GHz}$ and $\nu=500~\mathrm{GHz}$. The dashed lines represent the normalized contribution from Rayleigh scattering alone. The Rayleigh scattering signal shifts the peak of the visibility function towards lower redshifts.}
\label{visibility}
\end{figure}

However, the recombination processes also yield the production of neutral species, mainly hydrogen and helium, by which CMB photons can be scattered through Rayleigh scattering. Unlike Thomson scattering, the Rayleigh scattering cross section is frequency dependent: 
\begin{equation}
  \sigma_R(\nu)=\sigma_T \sum_{n=1} a_n\left( \frac{\nu}{\nu_0}\right)^{2n+2},
\end{equation}
where $a_1 = 1$, $a_2 = 2.626$, $a_3 = 5.502$ \citep{Lewis2013}, $\sigma_T$ is the cross section of Thomson scattering and $\nu_0=3.102\times 10^6~\mathrm{GHz}$ is the Lyman limit frequency.
After recombination, neutral species quickly dilute due to the cosmic expansion (scaling as $\propto a^{-3}$) and photons are diluted and redshifted ($\propto a^{-4}$), making the probability of Rayleigh scattering $\propto a^{-7}$. As a result, most Rayleigh scattering events occur shortly after recombination. This can be seen in the visibility function $g(z) \equiv \dot{\tau}e^{-\tau}$ where $\tau$ is the optical depth. The visibility function describes the probability that a photon last scattered at a given redshift. The visibility functions are shown in Fig.~\ref{visibility} for both Thomson and Rayleigh scattering.

By increasing the comoving opacity of the plasma in a frequency-dependent way, Rayleigh scattering leaves characteristic imprints on the CMB: 
\begin{itemize}
\item {\bf Frequency-dependent shift of the visibility function:}
    As can be seen in Fig.~\ref{visibility}, the visibility function is shifted towards lower redshifts (later time). Crucially, the position of the peak of the visibility function is made frequency dependent with the shift towards later time being larger at higher frequencies. 
\item {\bf Increase of diffusion damping:}
    Rayleigh scattering globally increases the comoving opacity of the plasma. The mean free path of photons in the plasma is therefore reduced. The strength of diffusion damping is directly controlled by the mean free path~\citep{visibility}. At lower frequencies, Rayleigh scattering has negligible impact. However, at higher frequencies, the visibility function is shifted towards later time when cosmic expansion has led to a global decrease of the comoving opacity, leading to an overall increase of diffusion damping. 
\item {\bf Frequency-dependent sound horizon:}
    As the peak of the visibility function is rendered frequency dependent by Rayleigh scattering, the size of the sound horizon at last scattering similarly inherits the same frequency dependence, being larger at higher frequencies. The size of the sound horizon dictates the location of the acoustic peaks in both the matter and the CMB power spectra, a larger sound horizon shifts the acoustic peaks towards larger scales.
\item {\bf Frequency-dependent amplitude of polarization signal:} 
    The shift of the visibility function induced by Rayleigh scattering increases the amplitude of the local temperature quadrupole around the time of recombination, leading to a boost in $E$ modes on large scales.
\end{itemize}

\begin{figure*}[!ht]
\centering
 \subfloat{\includegraphics[width=0.49\textwidth]{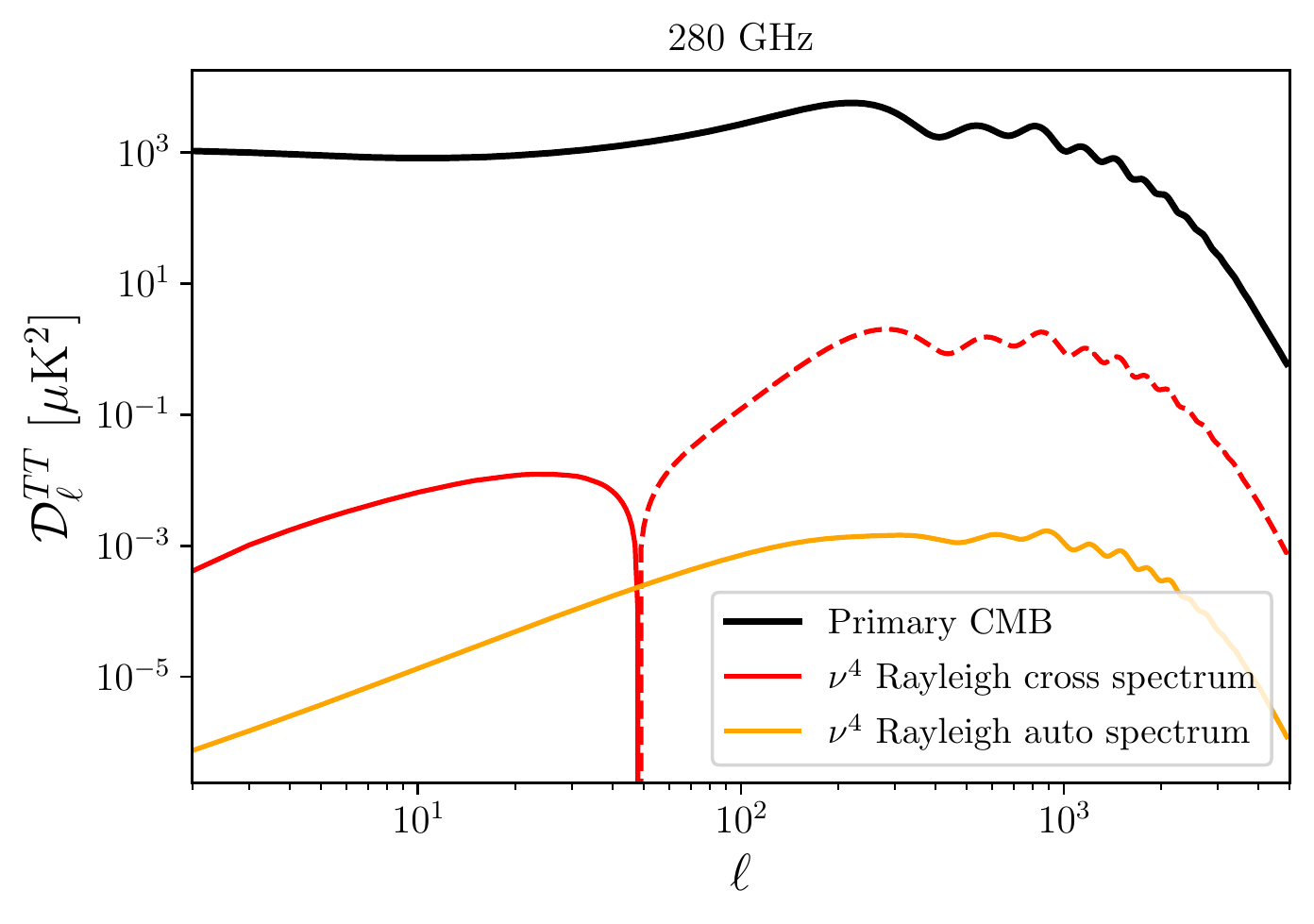}}
 \hfill
 \subfloat{\includegraphics[width=0.49\textwidth]{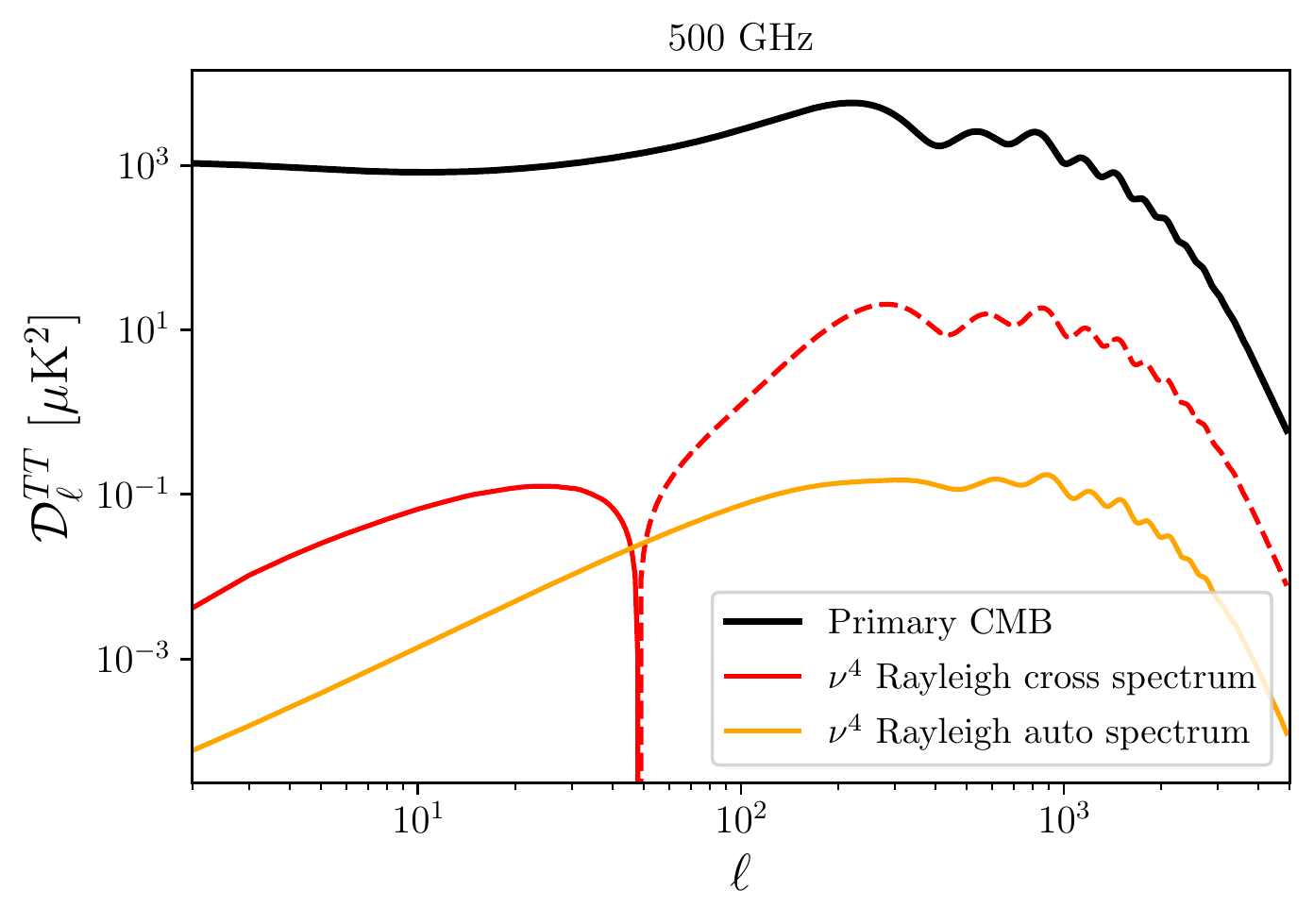}}
\caption{Contributions from Rayleigh scattering at $\nu=280~\mathrm{GHz}$ and $\nu=500~\mathrm{GHz}$. Red lines show the cross-spectra between the Rayleigh signal and the primary CMB. The solid lines present positive and the dotted parts are negative correlations. Yellow lines show the auto-spectra of the Rayleigh signal. The cross-spectra are roughly $3$ orders of magnitude larger than the auto-spectra of the Rayleigh signal.
}

\label{contribution}
\end{figure*}

If the shift of the visibility function is small enough, the effect of Rayleigh scattering on the temperature and polarization power spectra is accurately described by a linear, frequency-dependent distortion \citep{Lewis2013}:
\begin{equation}
    a_{\ell m}^{X}(\nu)=a_{\ell m}^{X}+\Big(\frac{\nu}{\nu_0}\Big)^4\Delta a_{\ell m}^{X,4}+ \cdots,
    \label{eq:rs-sht}
\end{equation}
where $a_{\ell m}$ are the coefficients in the spherical harmonic transform of the temperature ($X = T$) and polarization ($X = E,B$) anisotropies. Note that we have neglected higher order corrections to the Rayleigh scattering cross section that would give rise to additional terms on the right hand side of Eq.~\eqref{eq:rs-sht}. The power spectra are then given by
\begin{equation}
  \begin{aligned}
  C_{\ell}^{XY}(\nu_1,\nu_2)&=\left\langle a_{\ell m}^{{X}}(\nu_1)a_{\ell m}^{{Y}}(\nu_2) \right\rangle \\ &=C_{\ell}^{XY}+ \left(\frac{1}{\nu_0}\right)^4 \left(\nu_1^4C_{\ell}^{X\Delta Y_4}+\nu_2^4C_{\ell}^{\Delta X_4Y}\right)\\
  &\quad+\left(\frac{\nu_1\nu_2}{\nu_0^2}\right)^4 C_{\ell}^{\Delta X_4\Delta Y_4}+\cdots .
  \end{aligned}
\end{equation}
The first term in this expansion is the primary CMB, sourced by Thomson scattering only. The following term contains the cross-correlation spectra between the primary CMB and the Rayleigh scattering signal. The last term is proportional to the Rayleigh scattering auto-spectrum. These terms are also ordered with decreasing amplitude (see Fig.~\ref{contribution}) making the cross-correlation the natural target to achieve a first detection.

With the low-noise and large-area maps expected from future CMB experiments, the detection of this faint Rayleigh scattering signal should be possible~\citep{Lewis2013,Beringue2020}. More realistic estimates of the expected detection significance require forecasts including foregrounds and testing the robustness of these predictions after applying foreground cleaning techniques. 

\section{Foreground Modeling}
\label{sec:fgmodel}
\begin{figure*}[!ht]
\centering
 \subfloat{\includegraphics[width=0.48\textwidth]{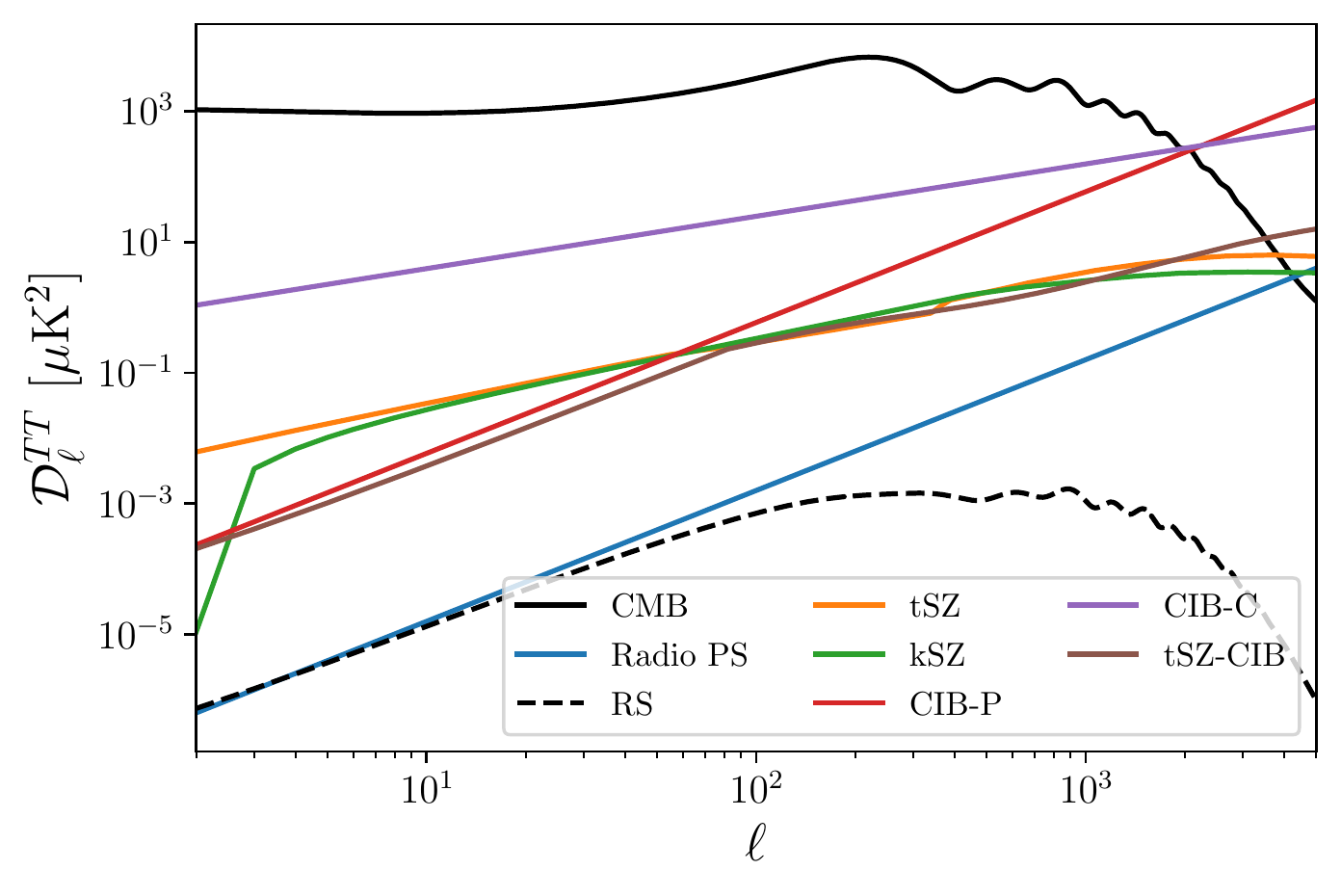}}
 \hfill
 \subfloat{\includegraphics[width=0.48\textwidth]{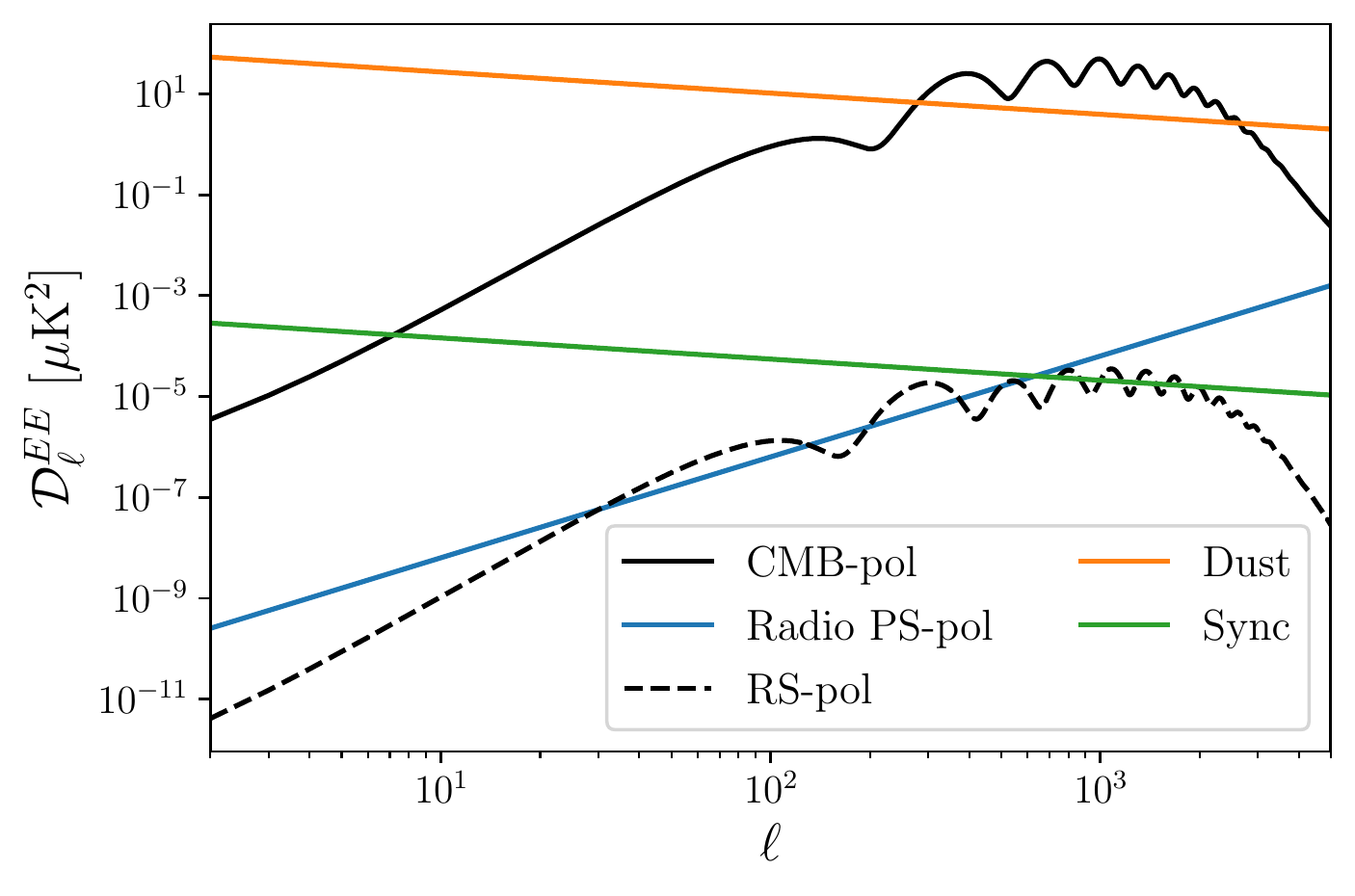}}
\caption{Signal and foreground power spectra at $\nu=280~\mathrm{GHz}$. The observed temperature includes contributions from the CMB primary signal, Rayleigh scattering auto signal, radio point sources, thermal Sunyaev–Zeldovich (tSZ), kinetic Sunyaev–Zeldovich (kSZ), cosmic infrared background (CIB), and cross correlation of tSZ and CIB; for polarization, the polarized primary CMB and polarized Rayleigh signal, dust, synchrotron, and radio point sources are included.}
\label{fg}
\end{figure*}

The microwave sky foreground models we considered in this paper include the cosmic infrared background (CIB), the thermal Sunyaev-Zeldovich effect (tSZ), the kinetic Sunyaev-Zeldovich (kSZ), radio sources, and both galactic dust and synchrotron emissions. The CIB is sourced by star-forming dusty galaxies and is modelled according to \citet{Dunkley2013}. The tSZ and kSZ are the two dominant effects in the scattering of CMB photons by free electrons in clusters. We model the tSZ and kSZ auto-spectra using templates from \citet{BBPSS}. We also include the cross-spectra between tSZ and CIB in our foreground modeling, following \citet{addison2012}. In principle there is also a polarized version of the SZ effect (pSZ) and the relativistic SZ effect (rSZ), but these are too small to have an impact on our forecasts, so we do not consider them in this analysis. Radio point sources (radio galaxies) are modeled following \citet{SOforecasts}. The angular power spectrum in temperature and polarization for all of these foregrounds, the CMB, and Rayleigh scattering are shown in Figure~\ref{fg}. 

The polarized dust and synchrotron foregrounds from our galaxies are modeled following \citet{DUSTSYNC}. For the dust model the emissivity is assumed to follow:
\begin{equation}
  \epsilon_\mathrm{d}(\nu,\beta_\mathrm{d})=\left(\frac{\nu}{\nu_0}\right)^{\beta_\mathrm{d}-2}\frac{B_{\nu}(T_\mathrm{d})}{B_{\nu_0}(T_\mathrm{d})}.
\end{equation}
Here $\beta_\mathrm{d}=1.59$, $T_\mathrm{d}=19.6~\mathrm{K}$, $\nu_0=353~\mathrm{GHz}$, and $B_{\nu}$ is the Planck function.

We can then express the dust and synchrotron power spectrum as:
\begin{align}
  \mathcal{D}^{\mathrm{d},\mathrm{s}}_{\ell}(\nu_1,\nu_2)   =& \left[\alpha_\mathrm{s}^2\left(\frac{\nu_1\nu_2}{\nu_K^2}\right)^{\beta_\mathrm{s}}+\alpha_\mathrm{d}^2\epsilon_\mathrm{d}(\nu_1,\beta_\mathrm{d})\epsilon_\mathrm{d}(\nu_2,\beta_\mathrm{d})\right] \nonumber \\
  &\times
  g(\nu_1)g(\nu_2)\left(\frac{\ell}{80}\right)^{-0.42} \, ,
\end{align}
where $\alpha_\mathrm{s}$ and $\alpha_\mathrm{d}$ are two scaling constants, which we approximate as $\alpha_\mathrm{s}^2=10.3~\mu \mathrm{K}^2$ and $\alpha_\mathrm{d}^2=0.53~\mu \mathrm{K}^2$ in antenna temperature, $\beta_\mathrm{s}=-3.1$ is the synchrotron spectral index. These parameters are measured from the cleanest 45\% of the sky, on angular scales with $\ell=50\mhyphen 110$ \citep[][]{DUSTSYNC}. We ignore the cross-spectra between dust and synchrotron since it is small \citep[][]{DUSTSYNC}. Finally, $g(\nu)$ is the conversion function from antenna to CMB thermodynamic temperature at frequency $\nu$:
\begin{equation}
  g(\nu)=\frac{\left(e^{\frac{h\nu}{k_\mathrm{B} T_\mathrm{CMB}}}-1\right)^2}{e^{\frac{h\nu}{k_\mathrm{B} T_\mathrm{CMB}}}\left(\frac{h\nu}{k_\mathrm{B} T_\mathrm{CMB}}\right)^2},
\end{equation}
where $h$ is the Planck constant and $k_\mathrm{B}$ is the Boltzmann constant.  

For radio point sources, we write the auto-spectrum as:
\begin{equation}
  \mathcal{D}_{\ell}^\mathrm{ps}(\nu_1,\nu_2)=A_\mathrm{ps}\left(\frac{\nu_1\nu_2}{\nu_r^2}\right)^{-2.5}\frac{g(\nu_1)g(\nu_2)}{(g(\nu_r))^2}\frac{\ell^2}{\ell_0^2},
  \label{eq:radio-ps}
\end{equation}
where $A_\mathrm{ps}=3.1~\mu \mathrm{K}^2$, $\ell_0=3000$, $\nu_r=150~\mathrm{GHz}$. 

For polarized radio point sources, we modify Eq.~\eqref{eq:radio-ps} as:
\begin{equation}
  \mathcal{D}_{\ell}^\mathrm{ps-pol}(\nu_1,\nu_2)=A_\mathrm{ps-pol}\left(\frac{\nu_1\nu_2}{\nu_r^2}\right)^{-2.5}\frac{g(\nu_1)g(\nu_2)}{(g(\nu_r))^2}\frac{\ell^2}{\ell_0^2}
\end{equation}
where $A_\mathrm{ps-pol}=0.015~\mu\mathrm{K}^2$ at $\ell_0=3000$ and $\nu_r=150~\mathrm{GHz}$\citep{SOforecasts}.
We assume that the galactic dust and synchrotron do not depend on the position of the sky observed. We show all auto-spectra at $\nu=280~\mathrm{GHz}$ in Fig.~\ref{fg} for temperature and polarization. We apply the best-fit value from \citet{Dunkley2013} to the extragalactic foreground models.

\section{Instrumental and Atmospheric Noise Models}
\label{sec:Tele}
We consistently model the noise power spectra of the surveys we considered, CCAT-prime, \textit{Planck}, Simons Observatory (SO), PICO and LiteBIRD using a model that consists in a {\it red} and a {\it white} component:
\begin{equation}
  n^\mathrm{ins}_{\ell}=n_\mathrm{red}\left(\frac{\ell}{\ell_\mathrm{knee}}\right)^{\alpha_\mathrm{knee}}+n_\mathrm{white} \, .
\end{equation}
Here $n_\mathrm{white}$ is the white noise due to the instrument alone, $n_\mathrm{red}$ is the red noise due to the atmosphere, where $\ell_\mathrm{knee}$ and $\alpha_\mathrm{knee}$ parameterize the scales affected by atmospheric contamination \citep{SOforecasts}. For satellite experiments such as {\it Planck}, we simply set $n_\mathrm{red}=0$. For ground-based observations, the red noise component is estimated from empirical measurements of the large angular scale noise from ACT at $150 ~{\rm GHz}$, and extrapolated to higher frequencies for SO, CCAT-prime, and CMB-S4 \citep{SOforecasts,CCATp,CMBS4}. Our forecasts will include \textit{Planck} to complement future ground-based experiments on the largest angular scales~\citep{Planck2018}.

The Simons Observatory consists in a set of telescopes under construction in Chile. Its large aperture telescope (LAT) will have a small beam allowing for precision measurement of small angular scales.  We refer to the SO LAT as `SO' throughout \citep{SOforecasts}. The Simons Observatory experiment also includes three small aperture telescopes (SATs), with larger beams. Because of the large beams, these SATs will be limited to $\ell<400$. With polarization modulators, SATs will only produce polarization maps; hence we do not consider temperature anisotropies measurements from the SATs.

The CCAT-prime observatory, currently under construction in Chile, will consist in the Fred Young Sub-millimeter Telescope (FYST) equipped with two instruments: an imaging camera, Prime-CAM \citep{Choi_2020} and a multi-beam submillimeter heterodyne spectrometer, CHAI \citep{CHAI}. We will refer to the FYST + Prime-CAM configuration as `CCAT-prime' for simplicity. Similar to SO, CCAT-prime will have a high sensitivity at small angular scales \citep{Choi_2020}. Compared to the SO LAT, CCAT-prime will focus on higher frequencies. Besides, the higher altitude of the site is expected to yield lower atmospheric noise.

LiteBIRD is a planned satellite mission covering a large frequency range. Due to the relatively small dish the beam will limit its resolution to $\ell < 400$~\citep{LiteBird}.

The Probe of Inflation and Cosmic Origin (PICO) is a recently proposed space satellite \cite{PICO}, with extremely low noise and many frequency channels. PICO will cover a large range of multipoles. If funded, it will be capable of a high significance measurement of Rayleigh scattering \citep{Beringue2020}.

All the noise parameters for the experiments mentioned above are listed in Appendix~\ref{sec:Telescopes}. For the ground-based experiments, we expect some differences between the measured red noise components and extrapolated values we use. Additionally we do not consider that the atmospheric red noise can be removed using proposed analysis techniques \citep{SPTinprep}. Thus, our forecast in this respect are conservative, since if such techniques are viable this would only improve the forecasts we present.

\begin{figure*}[!ht]
\centering
 \subfloat{\includegraphics[width=0.483\linewidth]{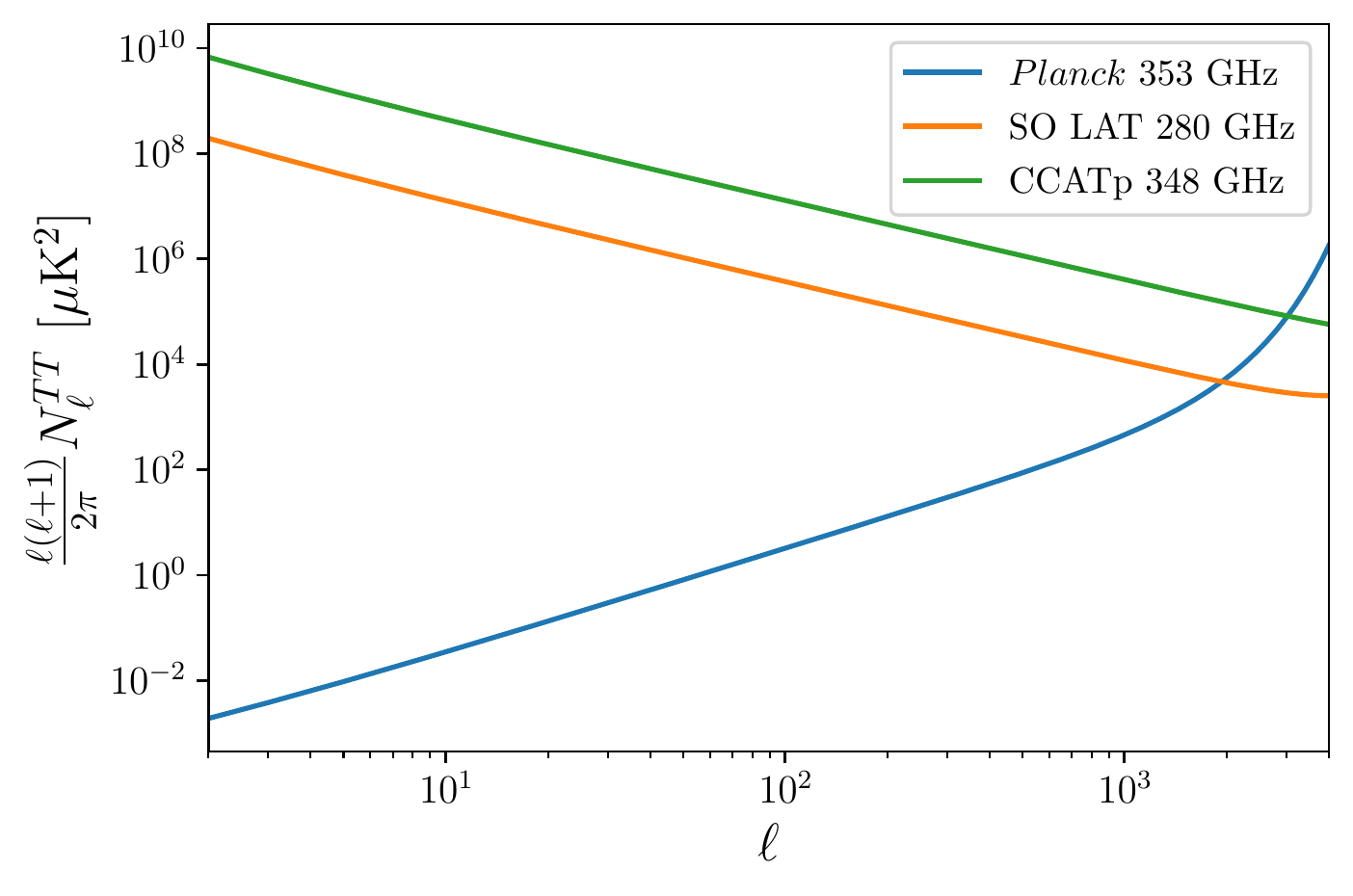}}
 \subfloat{\includegraphics[width=0.49\linewidth]{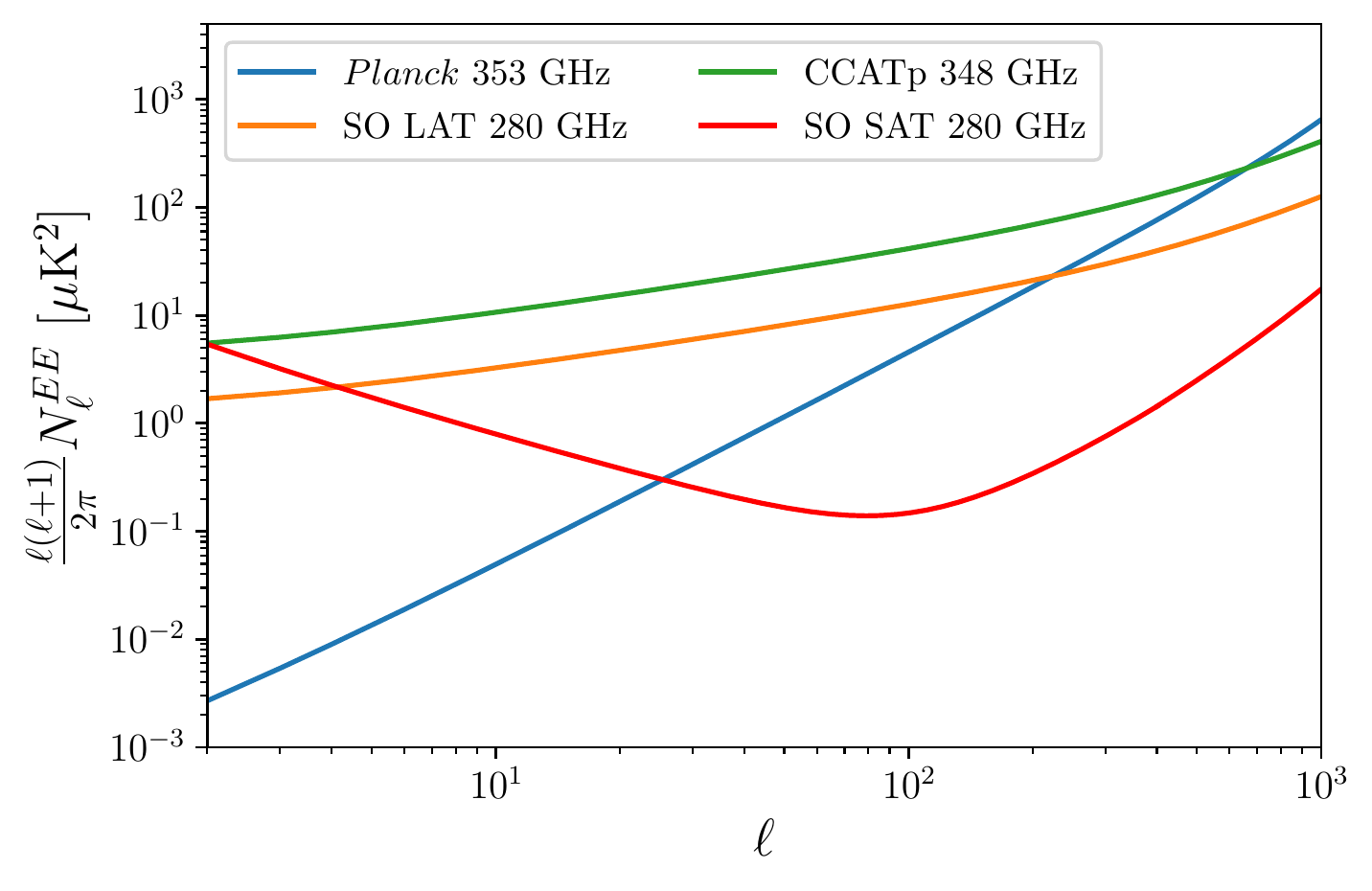}}
\caption{Comparison of instrumental noises of \textit{Planck}, CCAT-prime, SO LAT, and SO SAT at similar frequency channels. CCAT-prime and SO have smaller noise on small scales, while \textit{Planck} and SAT have smaller noise on large scales. In practice, we only use data from SO SAT for $\ell\leq400$.}

\label{fig:instrumental}
\end{figure*}

\section{Component Separation and Reconstructed Noise Spectra}
\label{sec:ILC}

The forecasts carried out in this work follow the constrained Internal Linear Combination (cILC) framework derived in \citet{Remazeilles_2010}\footnote{Technically, our analysis is not an cILC, since we are not constructing the covariance matrix from a map or data, but we will refer to it as such.}. Since we are interested in measuring the cross correlation between the primary CMB and the Rayleigh scattering signals, we schematically need to produce two maps: one that contains only the CMB signal and no Rayleigh scattering and another one that contains only the Rayleigh scattering signal and no primary CMB. This procedure will ensure that no residual CMB contamination in the Rayleigh scattering map (and vice-versa) would bias the cross correlation. We model the total data vector at a pixel $p$ as:
\begin{equation}
\mathbf{d}(p) = \mathbf{a}^{\rm CMB} s^{\rm CMB}(p) + \mathbf{b}^{\rm RS} s^{\rm RS}(p) + \mathbf{n}(p),
\label{eq:signal}
\end{equation}
where $\mathbf{a}$ is the frequency dependence of the CMB (assumed to be unity at all frequencies), $\mathbf{b}$ is the frequency dependence of the Rayleigh scattering signal ($\propto \nu^4$). $\mathbf{n}$ captures the effect of foregrounds, instrumental and atmospheric noises. A pixel here refers to either an actual pixel in a sky map or a set of harmonic coefficients $(\ell,m)$ of the spherical harmonic transform.

The cILC procedure yields a set of weights $\mathbf{w}$ that, for the CMB map, satisfy:
\begin{itemize}
    \item $\mathbf{w}^\trans{}\cdot\mathbf{a} = 1$.
    \item $\mathbf{w}^\trans{}\cdot\mathbf{b} = 0$.
    \item $\hat{s}^{\rm CMB}(p) \equiv \mathbf{w}^\trans{}\cdot\mathbf{d}(p)$ has minimum variance.
\end{itemize}
These weights are given by \citep{Remazeilles_2010}: 
\begin{equation}
\mathbf{w_{a=1,b=0}}^\trans{} = \frac{ \left(\mathbf{b}^\trans{} \mathbf{\Tilde{R}_d}^{-1}   \mathbf{b}\right)\mathbf{a}^\trans{}  \mathbf{\Tilde{R}_d}^{-1} - \left(\mathbf{a}^\trans{} \mathbf{\Tilde{R}_d}^{-1}   \mathbf{b}\right)\mathbf{b}^\trans{}  \mathbf{\Tilde{R}_d}^{-1} }{ \left(\mathbf{a}^\trans{} \mathbf{\Tilde{R}_d}^{-1}   \mathbf{a}\right) \left(\mathbf{b}^\trans{} \mathbf{\Tilde{R}_d}^{-1}   \mathbf{b}\right) - \left(\mathbf{a}^\trans{} \mathbf{\Tilde{R}_d}^{-1}   \mathbf{b}\right)^2}.
\label{eq:cilc-weights}
\end{equation}
Here $\mathbf{\Tilde{R}_d}$ is the covariance of the input maps. If the cILC is carried out in the harmonic domain, the weights will inherit a scale dependence: $\mathbf{w} \longrightarrow \mathbf{w}_\ell$. Finally, we note that the same procedure yields the weights for the Rayleigh scattering map ($\mathbf{w_{a=0,b=1}}$) by simply inverting $\mathbf{a}$ and $\mathbf{b}$ in Eq.~\eqref{eq:cilc-weights}.

From the linear combinations $\hat{s}^{\rm CMB}(p) = \mathbf{w_{a=1,b=0}}^\trans{} \cdot \mathbf{d}(p)$ and $\hat{s}^{\rm RS}(p) = \mathbf{w_{a=0,b=1}}^\trans{} \cdot \mathbf{d}(p)$ we can compute the primary CMB and Rayleigh scattering auto and cross-spectra, that, given the constraints obeyed by the weights, read:
\begin{align}
    \hat{C}_\ell^{\rm CMB} &= C_\ell^{\rm CMB} + \mathbf{w_{a=1,b=0}}^\trans{}  \mathbf{N}_\ell   \mathbf{w_{a=1,b=0}}, \nonumber \\
    &= C_\ell^{\rm CMB} + N_\ell^{\rm CMB}, \nonumber \\
    \hat{C}_\ell^{\rm RS} &= C_\ell^{\rm RS} + \mathbf{w_{a=0,b=1}}^\trans{}  \mathbf{N}_\ell   \mathbf{w_{a=0,b=1}}, \nonumber \\
    &= C_\ell^{\rm RS} + N_\ell^{\rm RS}, \nonumber \\
    \hat{C}_\ell^{\rm CMB \times RS} &= C_\ell^{\rm CMB\times RS} + \mathbf{w_{a=0,b=1}}^\trans{}  \mathbf{N}_\ell   \mathbf{w_{a=1,b=0}}, \nonumber \\
    &= C_\ell^{\rm CMB\times RS} + N_\ell^{\rm CMB \times RS}.
\end{align}
Here $\mathbf{N}_\ell$ is the covariance of the foregrounds, instrumental and atmospheric noises. We also note that when we consider the cross-correlation between the primary temperature CMB and the Rayleigh scattering $E$-mode polarization signal (or vice-versa), the noise term $N_\ell^{\rm CMB \times RS}$ vanishes due to our modelling of foregrounds and noise.

The variance of the reconstructed cross-spectrum is given by\citep{Knox_1995}, 
\begin{align}
   \Delta C_{\ell} &= \frac{1}{\sqrt{f_\mathrm{sky}(2\ell+1)}}\Big[ \left(C_{\ell}^\mathrm{RS}+N_{\ell}^\mathrm{RS}\right)
  \left(C_{\ell}^\mathrm{CMB}+N_{\ell}^\mathrm{CMB}\right)\nonumber \\
  & +\left(C_{\ell}^{\mathrm{CMB} \times \mathrm{RS}} + N_\ell^{\rm CMB \times RS}\right)^2\Big]^{1/2}.
\end{align}
The calculations above pertain to a given experiment. However, we are interested in the combination of experiments to maximize our ability to measure the Rayleigh scattering signal. To obtain a combined noise on the cross-spectra for a given set of experiments, we augment the data vector $\mathbf{d}(p)$ by taking the set of frequency channels to be:
\begin{equation}
    f^\mathrm{total}=(f^1,f^2,\cdots,f^i).
\end{equation}

The signal-to-noise of the cross-correlation is given by:
\begin{equation}
  \mathrm{SNR}^2
  =\sum_\ell\frac{\left(C_{\ell}^{\mathrm{CMB} \times \mathrm{RS}}\right)^2}{\Delta C_{\ell}^2}
\label{knox}
\end{equation}

\section{Results}
\label{sec:Results}

\begin{figure}[tb]

  \centering

  \includegraphics[width=\linewidth]{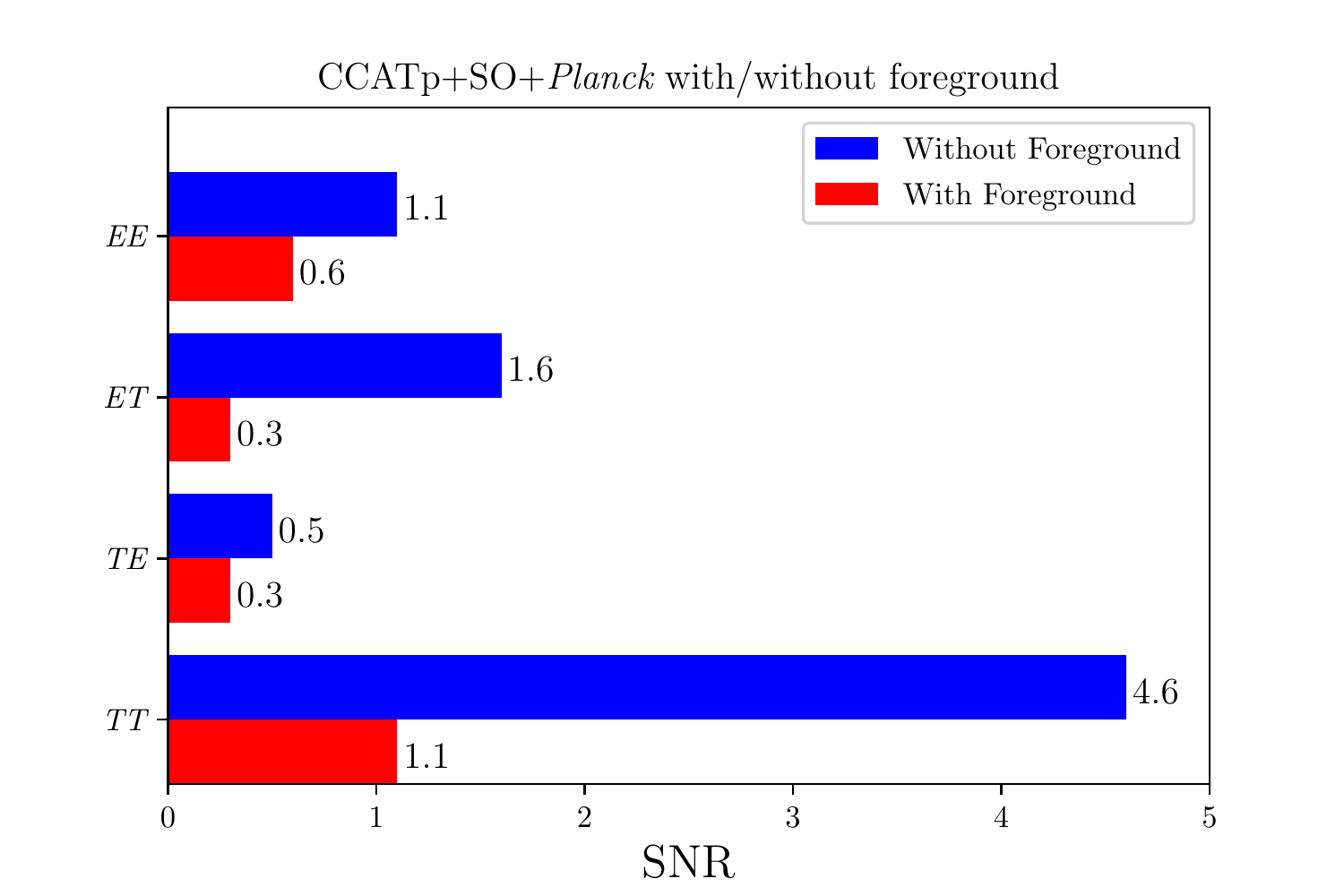}
 
\caption{Forecast of SNR with and without foregrounds for CMB Rayleigh scattering cross-spectrum using CCATp+SO+\textit{Planck}. The $TT$ SNR is significantly impacted by the foregrounds, being reduced by $77\%$ when foregrounds are present, which demonstrates that foregrounds are an hindrance for the first detection of the Rayleigh scattering signal.}
\label{fghitsn}
\end{figure}

The detectability for different experiments and combinations of experiments are summarized in Tab.~\ref{table}. In Fig.~\ref{fghitsn} we focus on the combination of CCAT-prime, SO and \textit{Planck} and show the SNR for the four CMB-Rayleigh scattering cross-spectra with and without foregrounds. For this combined configuration we estimate SNR of $1.1\sigma$ for the $TT$ signal with foregrounds present. For all other combinations, we find signal-to-noise below $1\sigma$. Without foregrounds, the combined configuration of CCAT-prime, SO and \textit{Planck} could reach $4.6\sigma$ in $TT$, $1.1\sigma$ in $EE$, and $1.6\sigma$ in $ET$. This clearly demonstrates the significant impact of foregrounds; for example the $TT$ SNR is reduced by $77\%$ by the inclusion of foregrounds and their removal. Similarly, the signal-to-noise in $ET$  reduces by $81\%$, $EE$ reduces by $45\%$ and $TE$ reduces by $40\%$. The latter is a result of the fact that in $TE$, foregrounds are not very significant.

\begin{table}[tb]
\begin{tabular}{|l|cccc|}
\hline
\rowcolor[HTML]{E6E8FF} 
                                                                       & \multicolumn{4}{c|}{\cellcolor[HTML]{E6E8FF}\textbf{SNR}}                                                                                                        \\ \hline
\rowcolor[HTML]{E6E8FF} 
\textbf{Configuration}                                                 & \multicolumn{1}{c|}{\cellcolor[HTML]{E6E8FF}$TT$} & \multicolumn{1}{c|}{\cellcolor[HTML]{E6E8FF}$TE$} & \multicolumn{1}{c|}{\cellcolor[HTML]{E6E8FF}$ET$} & $EE$ \\ \hline
\cellcolor[HTML]{E6E8FF}CCATp+SO+\textit{Planck}                          & \multicolumn{1}{c|}{1.1}                          & \multicolumn{1}{c|}{0.3}                         & \multicolumn{1}{c|}{0.3}                         & 0.6 \\ \hline
\rowcolor[HTML]{EFEFEF} 
\cellcolor[HTML]{E6E8FF}CCATp+SO+\textit{Planck} no Foreground                             & \multicolumn{1}{c|}{\cellcolor[HTML]{EFEFEF}4.6}  & \multicolumn{1}{c|}{\cellcolor[HTML]{EFEFEF}0.5} & \multicolumn{1}{c|}{\cellcolor[HTML]{EFEFEF}1.6}  & 1.1  \\ \hline
\cellcolor[HTML]{E6E8FF}LiteBIRD                                       & \multicolumn{1}{c|}{1.9}                          & \multicolumn{1}{c|}{0.1}                         & \multicolumn{1}{c|}{0.9}                         & 0.2 \\ \hline
\rowcolor[HTML]{EFEFEF} 
\cellcolor[HTML]{E6E8FF}LiteBIRD+CCATp+\textit{Planck} & \multicolumn{1}{c|}{\cellcolor[HTML]{EFEFEF}2.2}  & \multicolumn{1}{c|}{\cellcolor[HTML]{EFEFEF}0.2} & \multicolumn{1}{c|}{\cellcolor[HTML]{EFEFEF}0.9} & 0.4 \\ \hline
\cellcolor[HTML]{E6E8FF}PICO                                        & \multicolumn{1}{c|}{85}                          & \multicolumn{1}{c|}{17}                           & \multicolumn{1}{c|}{43}                           & 26   \\ \hline
\rowcolor[HTML]{EFEFEF} 
\end{tabular}
\caption{Forecasted signal-to-noise ratio for various configurations. The CCATp+SO+\textit{Planck} configuration can potentially achieve a $1.1\sigma$ detection with foreground present. We also show the impact of foregrounds for the combination CCATp+SO+\textit{Planck}. PICO, with its broad frequency coverage, is capable of a high-significance detection of Rayleigh scattering even after foreground removal.}
\label{table}
\end{table}

The SNR without foregrounds is calculated using a modified covariance matrix with only instrumental and atmospheric noise. As a consistency check we compare our SNR without foregrounds to the Fisher forecasts from \citet{Beringue2020}, which did not include any foregrounds. We find good agreements between the two methods, with the SNR from the cILC being slightly smaller due to the increase in variance caused by the deprojections.

\begin{figure*}[ht]
\centering
 \subfloat{\includegraphics[width=0.49\linewidth]{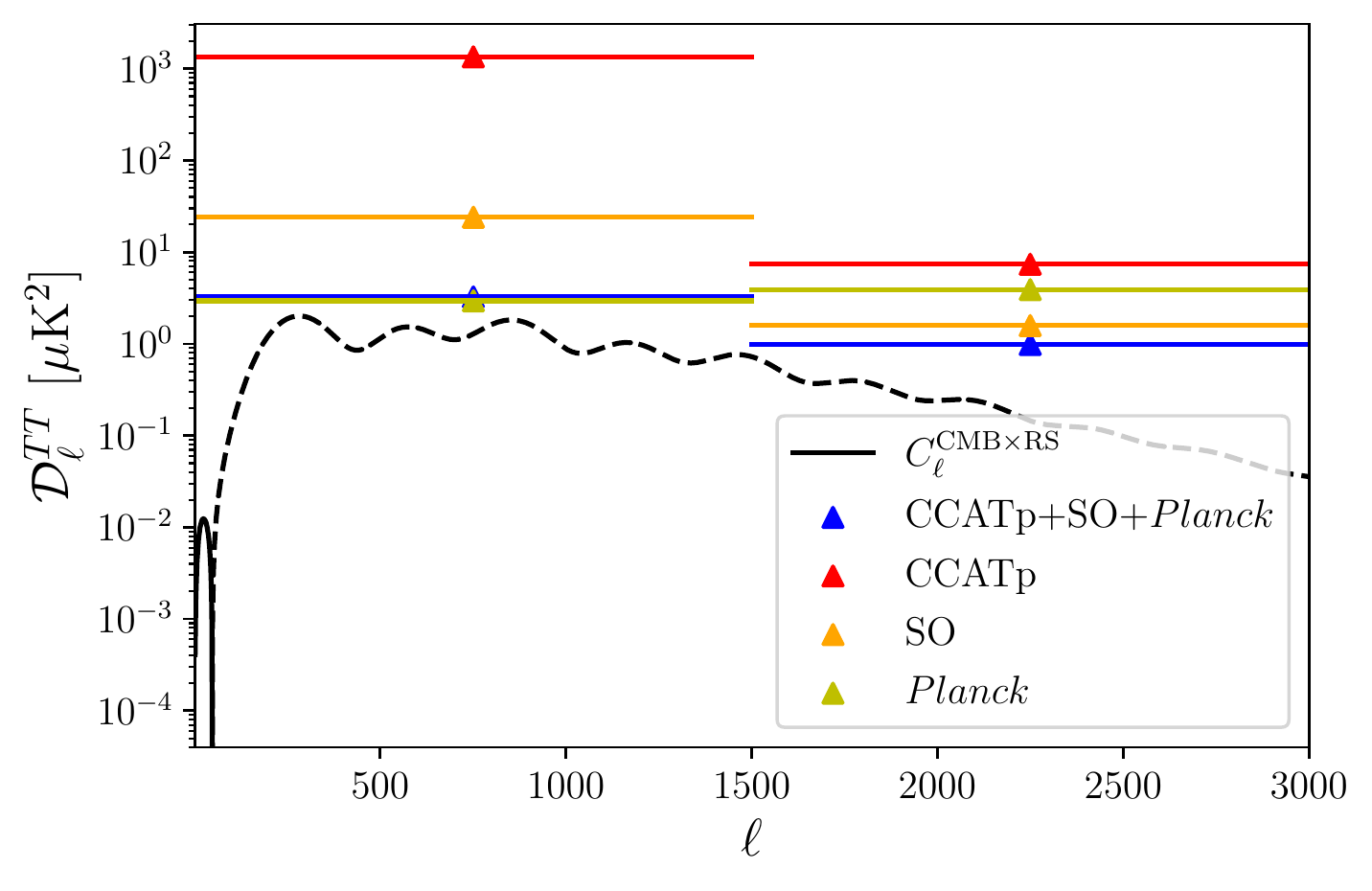}}
 \hfill
 \subfloat{\includegraphics[width=0.49\textwidth]{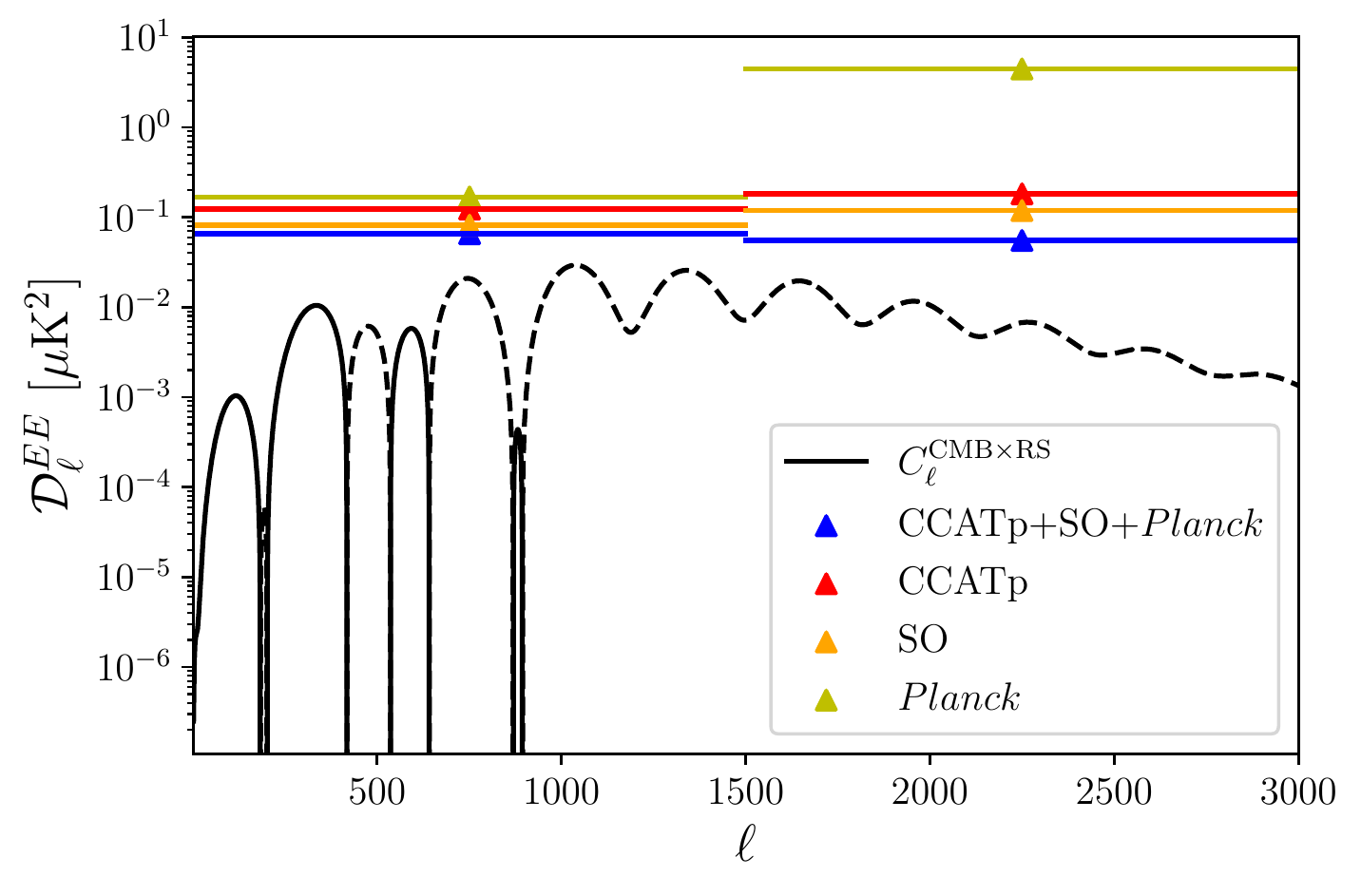}}

\caption{Forecasted $2\sigma$ upper limits on the Rayleigh scattering cross-spectrum with CCAT-prime, Simons Observatory LAT+SAT, \textit{Planck}, and the combination of all of these telescopes. The SO SAT only enters for the $EE$ signal, since it only targets polarized signals. As expected, the combined configuration has the lowest $2\sigma$ upper limits; \textit{Planck} is most helpful on large scales, and CCAT-prime and SO contribute mainly on small scales.}
\label{noisecomp}
\end{figure*}

In Fig.~\ref{noisecomp} we show the $2\sigma$ upper limit on the detectability of the $TT$ and $EE$ Rayleigh scattering signals for the CCAT-prime {\it Planck}, SO-LAT, and SO-SAT ($EE$ only) as well as their combination. As expected we find {\it Planck} to dominate the constraints on large scales in temperature as ground-based experiments are heavily impacted by the atmosphere. On small scales and in polarization, the improvements in both sensitivity and resolution enabled by ground-based surveys lead to SO and CCAT-prime driving the constraints.

We also estimate how much CCAT-prime is impacted by the atmosphere. The baseline signal-to-noise including atmosphere for $TT$ signal is $\mathrm{SNR}=0.07$, white it is $\mathrm{SNR}=1.79$ without atmosphere. If we combine the atmosphere-less CCAT-prime with SO and \textit{Planck}, we obtain $\mathrm{SNR}=1.88$ for the $TT$ channel. This suggests that the atmosphere severely limits detectability prospects for ground-based observations, thus mitigating atmospheric contributions plays an important role in the detection of Rayleigh scattering from the ground \citep{SPTinprep}. 

\begin{figure*}[!ht]
\centering
 \subfloat{\includegraphics[width=.49\linewidth]{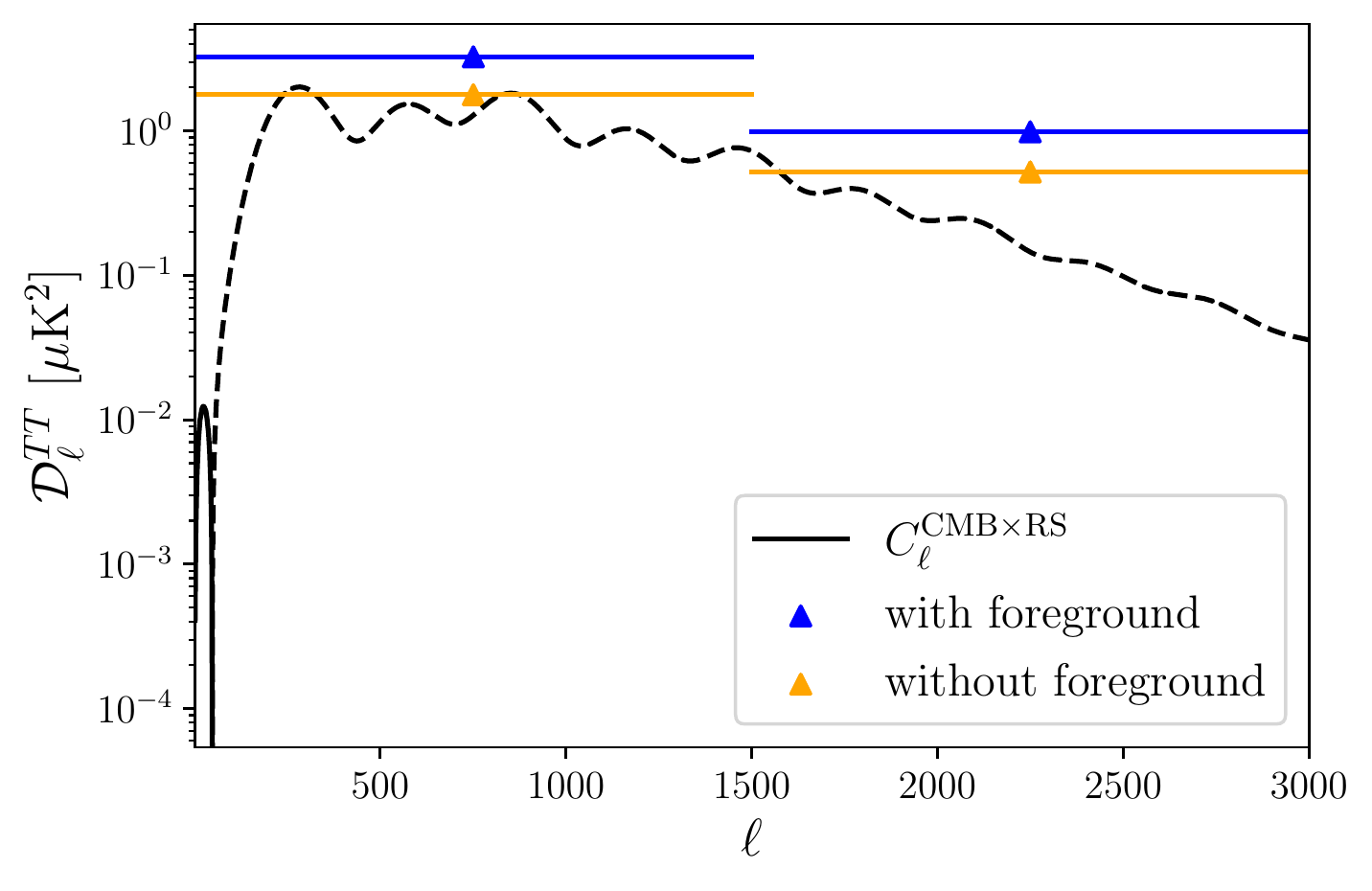}}
 \quad
 \subfloat{\includegraphics[width=.49\linewidth]{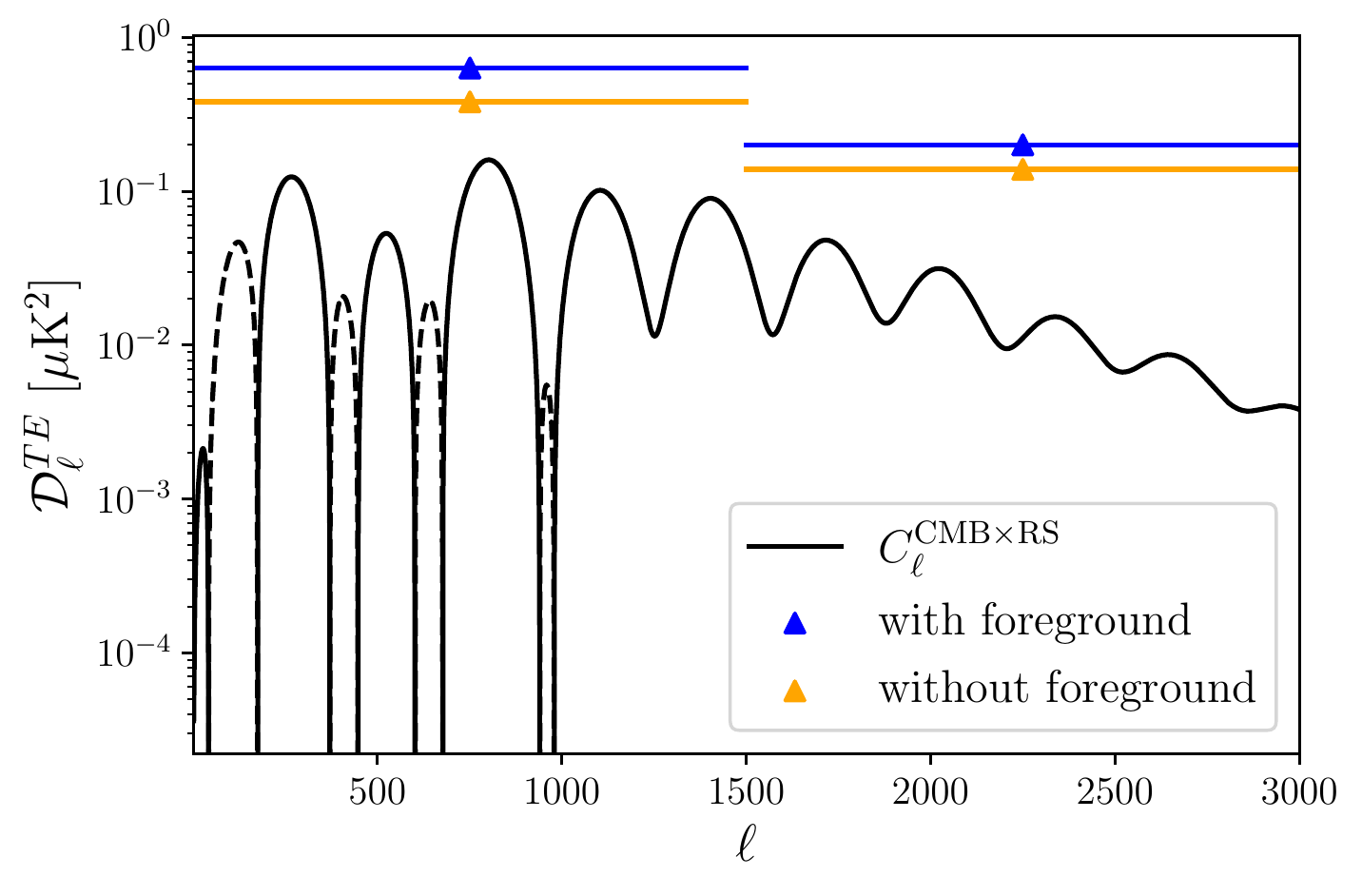}}
 \quad
 \subfloat{\includegraphics[width=.49\linewidth]{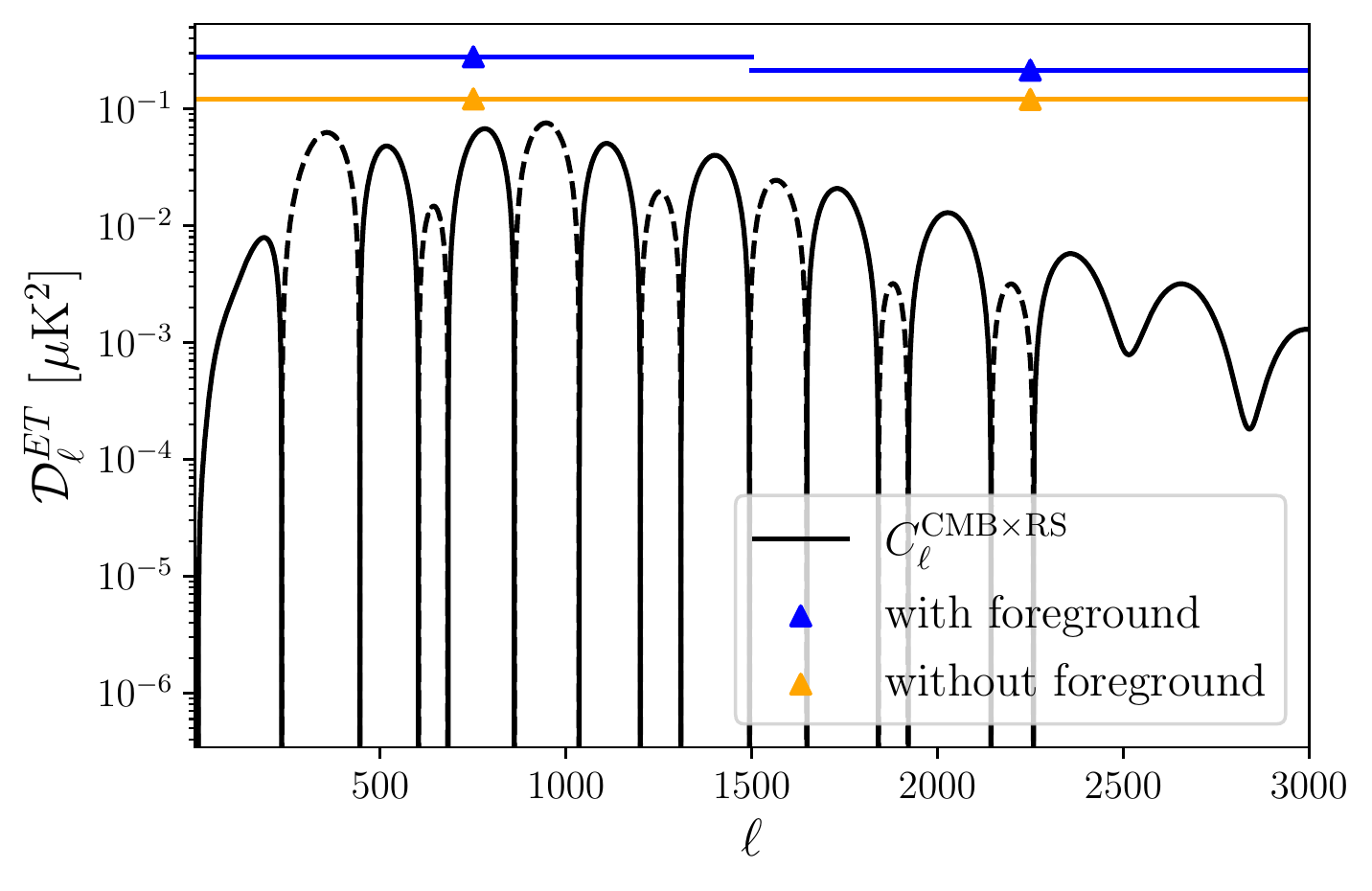}}
 \quad
 \subfloat{\includegraphics[width=.49\linewidth]{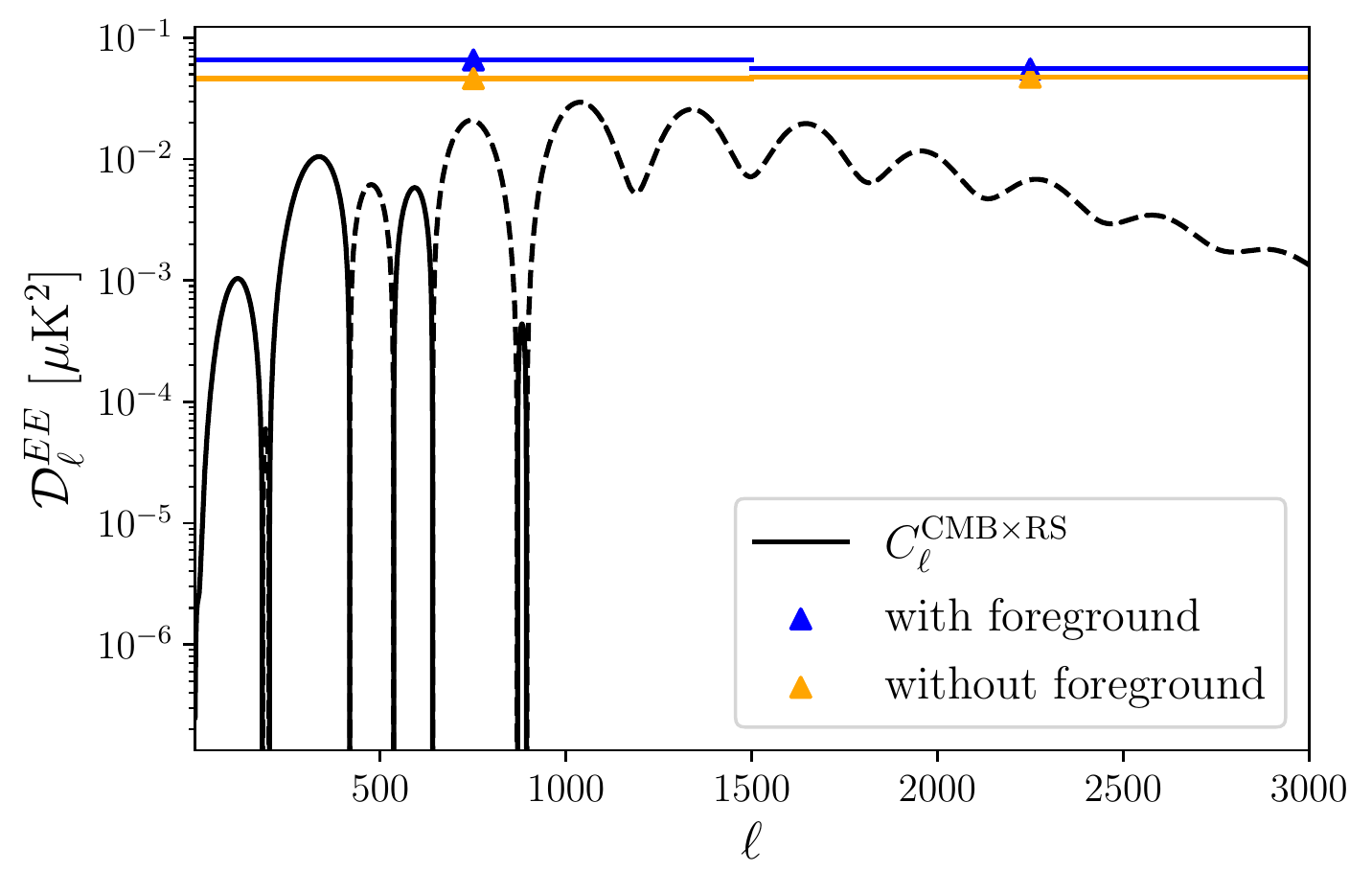}}
\caption{Forecasted $2\sigma$ upper limits on the $TT$ and $EE$ Rayleigh scattering cross-spectrum with and without foregrounds, demonstrating how the foregrounds impact detectability. The SO SAT only enters for the $EE$ signal, since it only targets polarized signals.}
\label{fghit}
\end{figure*}

\subsection{Residual Foreground Bias}
\label{sec:bias}

The cILC foreground cleaning method does not perfectly null all foregrounds, only those explicitly chosen to be nulled. As a result, residual signals from foreground not explicitly nulled remain and could possibly contaminate or bias the detection of Rayleigh scattering. For Rayleigh scattering this is especially important to check due to the relative amplitudes of foregrounds compared to Rayleigh cross-spectra. Knowing the amplitude of residuals will also inform whether further foreground cleaning is required. For this study we focus on temperature, since the foregrounds are less bright in polarization and hence less of a concern, as highlighted in Fig.~\ref{fghitsn}. 

We define the bias $B_{\ell}$ induced by residual foreground from a component with frequency-covariance matrix $\mathbf{S}_\ell$ to be: 
\begin{equation}
  B_{\ell}= \mathbf{w_{a=0,b=1}}^\trans{}  \mathbf{S}_{\ell}   \mathbf{w_{a=1,b=0}},
  \label{eq:bias}
\end{equation}
where $\mathbf{w_{a=0,b=1}}$ and $\mathbf{w_{a=1,b=0}}$ are defined in Eq.~\ref{eq:cilc-weights}. We consider tSZ, CIB, and radio point sources for $\mathbf{S}_\ell$. We are especially interested in the CIB contamination since CIB becomes the dominant foreground signal at sub-millimeter wavelengths where the Rayleigh signal is stronger.

\begin{figure}[!ht]
\centering
    \includegraphics[width=\columnwidth]{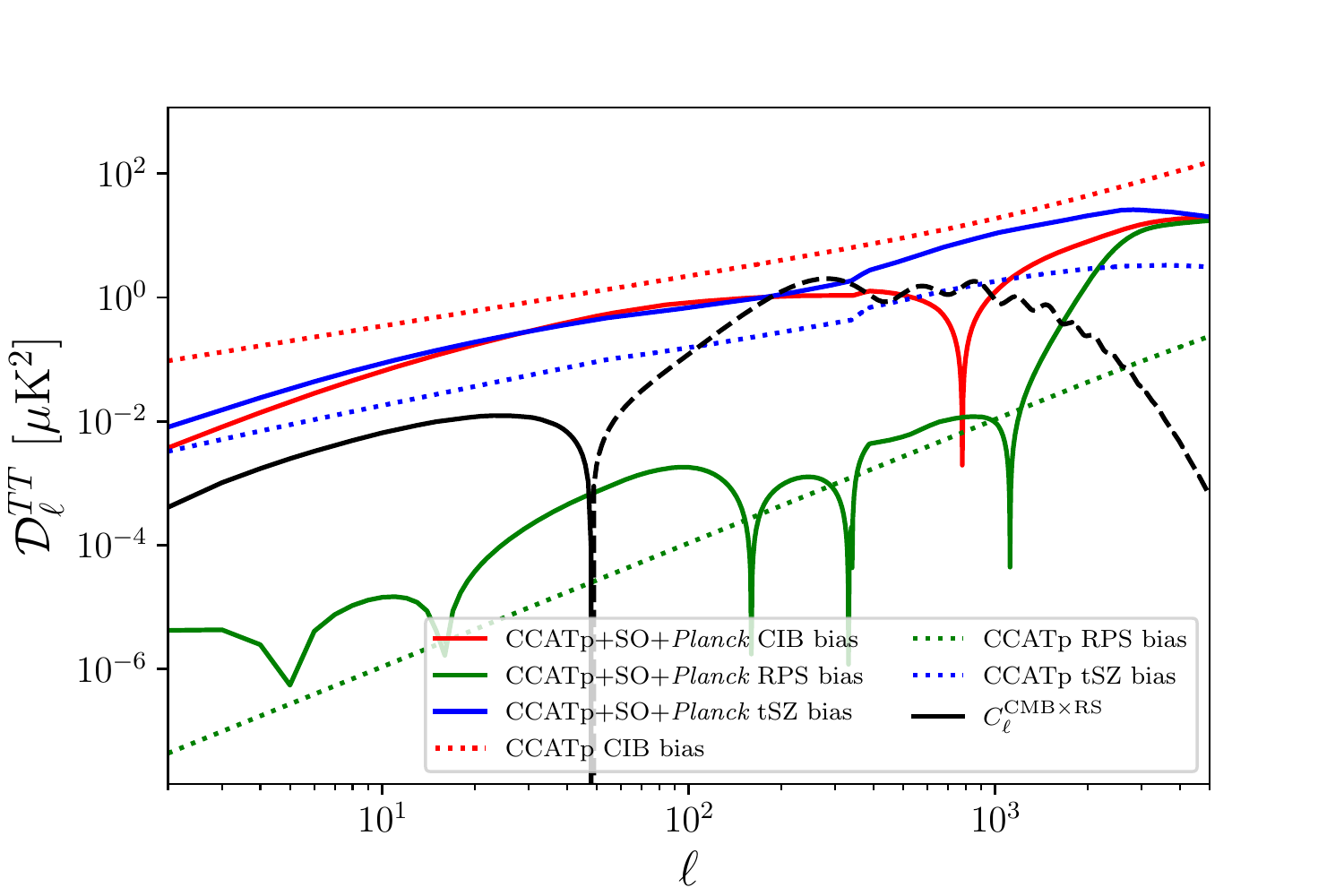}
 
\caption{CIB, tSZ, and radio point source bias for CCAT-prime and for the combined configuration. With only CCAT-prime, CIB is $1$ to $2$ orders of magnitude stronger than the Rayleigh scattering cross-spectrum at $280~\mathrm{GHz}$, and radio point source bias is around $1$ to $4$ orders of magnitude lower than the signal. The tSZ bias is comparable to the signal itself. With the combined configuration of CCAT-prime, SO and \textit{Planck}, we the CIB is more suppressed, but tSZ is less suppressed. In general, the tSZ and CIB biases are mostly comparable or even larger than the cross Rayleigh scattering cross-spectrum at $280~\mathrm{GHz}$.}
\label{bias}
\end{figure}

In Fig.~\ref{bias} we show the level of residual bias ($B_{\ell}$) resulting from tSZ, CIB, and radio point sources foregrounds calculated for both CCAT-prime and a combination of CCAT-prime, SO and {\it Planck}. Fig.~\ref{bias} illustrates that for CCAT-prime alone the CIB and tSZ are $1$ to $2$ orders of magnitude stronger than the CMB-Rayleigh cross-spectrum at $280~\mathrm{GHz}$, while radio point source residuals are around $1$ to $4$ orders of magnitude lower than the signal. Between $\ell=2000-3000$, we can see that the tSZ bias is slightly lower than the signal at $280~\mathrm{GHz}$.

Combining the experiments adds more frequency coverage, and the CIB residual is more suppressed, but the tSZ residual is less well suppressed. Given the results found here, ground-based detection of Rayleigh scattering is further complicated by CIB and tSZ residuals after the cILC cleaning. This necessitates the need to find methods to better constrain those components in the future or further constrain them in the cILC at the cost of SNR. 

\subsection{Experimental Calibration Tolerance}
\label{sec:gain}

Another potential source of hindrance to a first detection of Rayleigh scattering are instrumental systematic effects. In particular, we study the impact of imperfect calibration of a given experiment. We will refer to this overall calibration systematic as the gain. We model this gain systematic following \citet{Abitbol_2021}. Such gain calibration systematics might arise from fabrication variation in the on-chip bandpass filters between both detectors and wafers, time variation from atmospheric fluctuations, and systematic effects in the bandpass calibrations from measurements using Fourier transform spectrometers or other techniques. A formal model for gain calibration systematics would include an effective shift in frequency $\Delta\nu$, and a gain difference $\Delta g$, which describes how much of the sky emission signal the extra gain picks up \citep{Abitbol_2021}. We estimate an approximately $\Delta g=1\%$ calibration gain of the sky emission and do not consider the possible effective shift in the mean frequency to simplify our analysis.  We can express the gain as
\begin{equation}
  G_{\ell}=\Delta g\;\mathbf{w_{a=0,b=1}}^\trans{}  \mathbf{E}_{\ell}   \mathbf{w_{a=1,b=0}},
\end{equation}
where the sky emission $\mathbf{E}=\mathbf{S_{fg}}+\mathbf{S}$. Here $\mathbf{S_{fg}}$ represents the foregrounds, and $\mathbf{S}$ represents the signals.

\begin{figure*}[!ht]
\centering
 \subfloat{\includegraphics[width=0.49\linewidth]{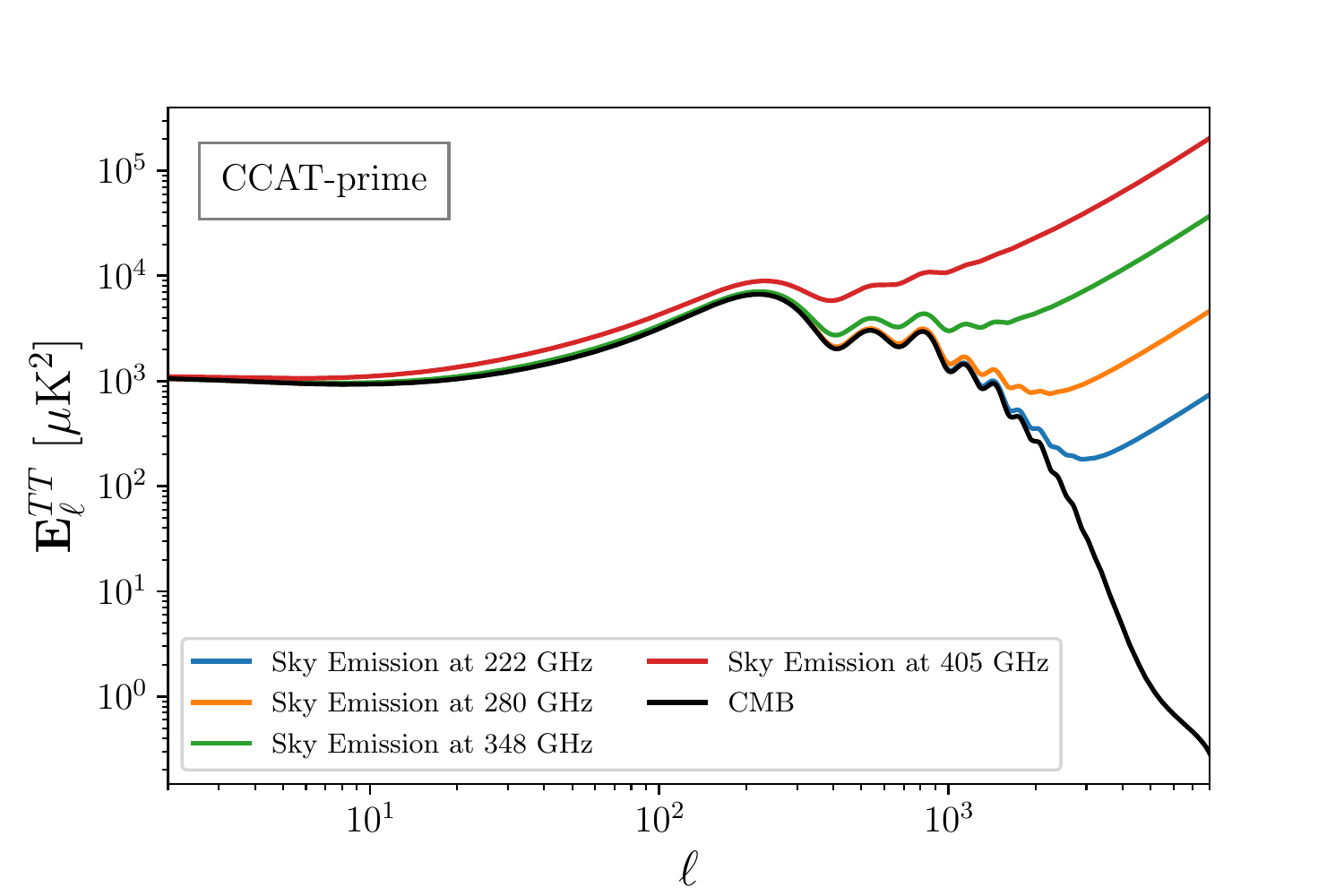}}
 \hfill
 \subfloat{\includegraphics[width=0.49\linewidth]{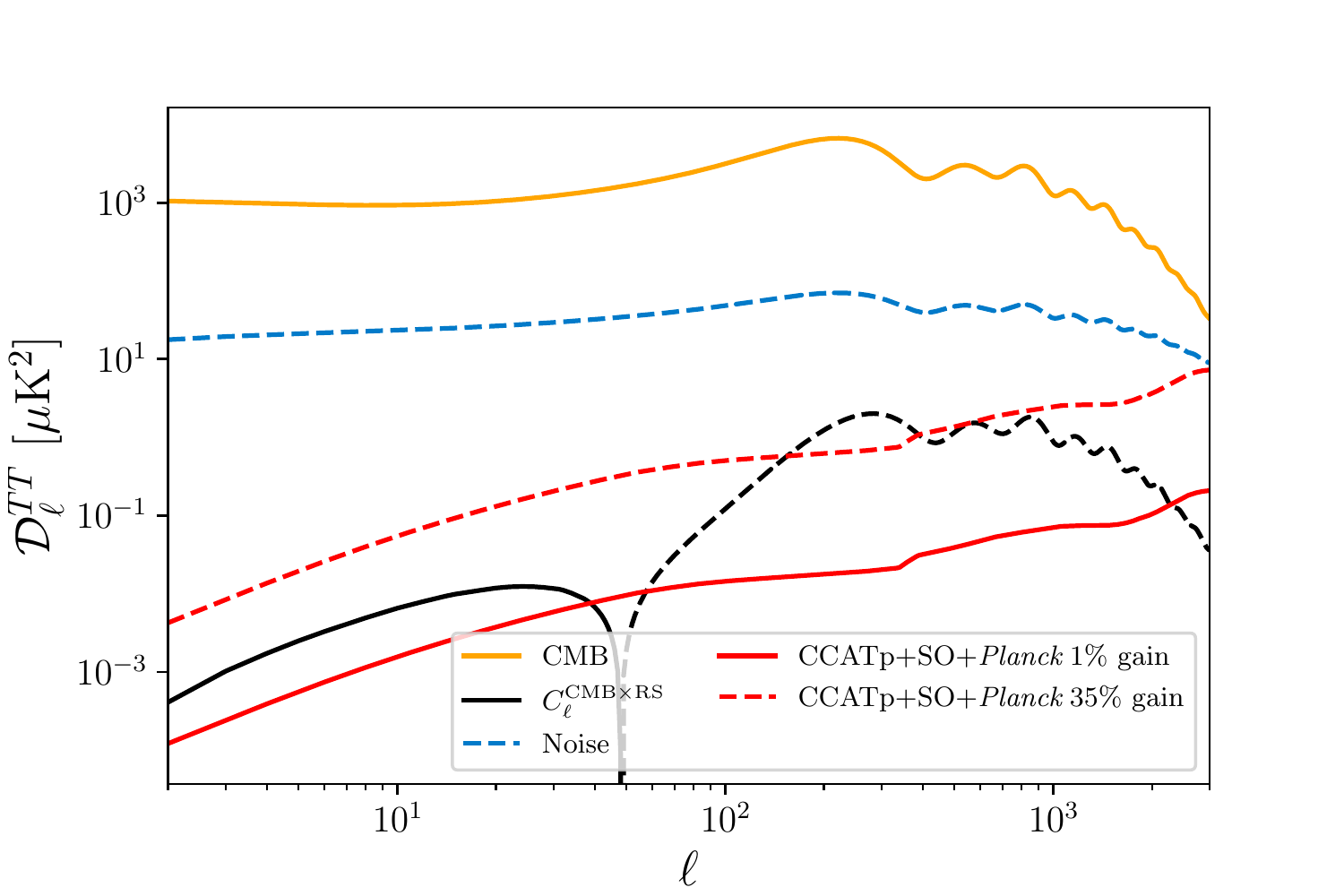}}
\caption{The left panel shows the sky emission on each channel of CCAT-prime. We omitted the $850~\mathrm{GHz}$ channel because its noise is so high that it is given a nearly zero weight by our cILC. We can see that at low frequencies, the sky emission is close to the CMB itself, but in higher frequency channels, the foreground terms start to dominate. The right panel is the gain model for the combined configuration of CCAT-prime, SO, and \textit{Planck}. We assumed $\Delta g=1\%$ gain from sky emission, and we can see that it is roughly on the same scale as the Rayleigh signal. With CCAT-prime channels, in lower frequency channels, the gain looks like CMB itself, but in the higher frequency channels, the gain gradually flattens out due to the dominance of radio point sources. The gain for the combined configuration is approximately $3$ to $4$ orders of magnitude lower than the noise power spectrum, meaning a $1\%$ gain calibration would be sufficient for our purpose. We also showed that a $35\%$ gain calibration, in dashed red line, would be comparable to the noise power spectrum at high $\ell$.}
\label{gain}
\end{figure*}

We show the sky emission in temperature for each channel for CCAT-prime in the right panel of Fig.~\ref{gain}. For the CCAT-prime lower frequency channels the sky emission looks like a CMB auto-spectrum, but at higher frequencies the extragalactic foreground signals are more dominant at high-$\ell$, especially the CIB.

We calculate the gain bias for the combined configuration of CCAT-prime, SO, and \textit{Planck} with $\Delta g=1\%$ and $\Delta g=35\%$. These biases are shown in left panel of Fig.~\ref{gain}. For $\Delta g=1\%$ of the sky emission the gain bias is around $3$ to $4$ orders of magnitude smaller than the noise power spectrum and would therefore be a sufficient calibration of instrument for our purpose. This is, in principle, informative in terms of the order of magnitude of the gain. It is helpful to verify that the gain is generally $4$ to $6$ orders of magnitude lower than our noise power spectrum. If we raise the gain to $\Delta g=35\%$, we see that the cross over happens at around $\ell=3000$. This shows that if we want to set our $\ell_\mathrm{max}$ of our observation higher than $\ell=3000$, a calibration around $1\%$ would be sufficient. A more systematic study is required to establish gain calibration thresholds that contain 2 significant figures for CCAT-prime, SO, and Planck.

\subsection{The Detection of Rayleigh Scattering by Satellite Missions}

\label{sec:space}
Ground-based observations are heavily impacted by atmospheric noise. Given the large impact of atmosphere on the detectability of the Rayleigh scattering signal, ideally one would attempt such observations in space where there is no atmosphere and where large scale anisotropies can be efficiently recovered. We calculate the detectability of the Rayleigh scattering signal for the planned satellite mission LiteBIRD~\cite{LiteBird} and the proposed mission PICO~\cite{PICO}.

\begin{figure*}[!ht]
\centering
 \subfloat{\includegraphics[width=0.49\linewidth]{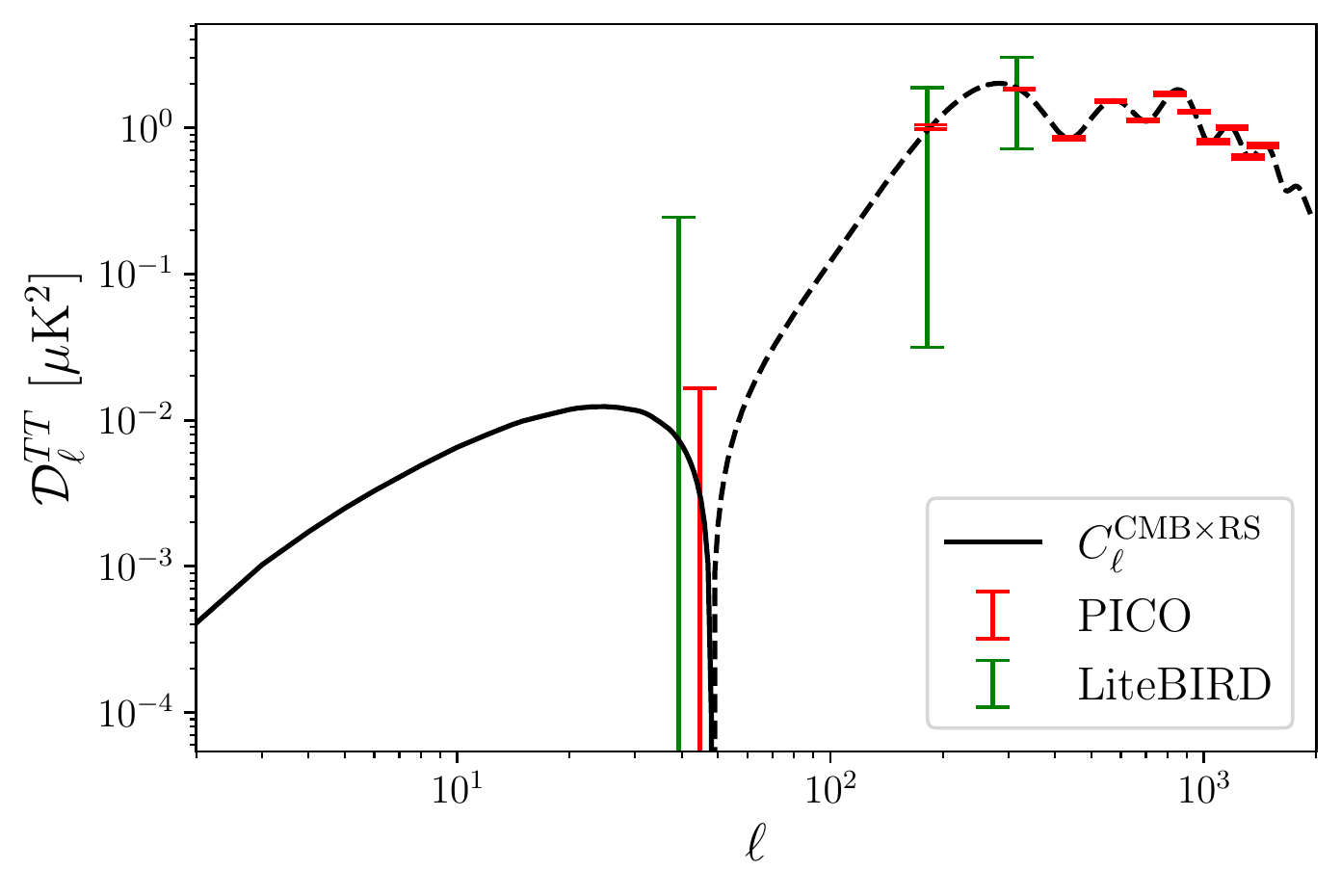}}
 \hfill
 \subfloat{\includegraphics[width=0.49\linewidth]{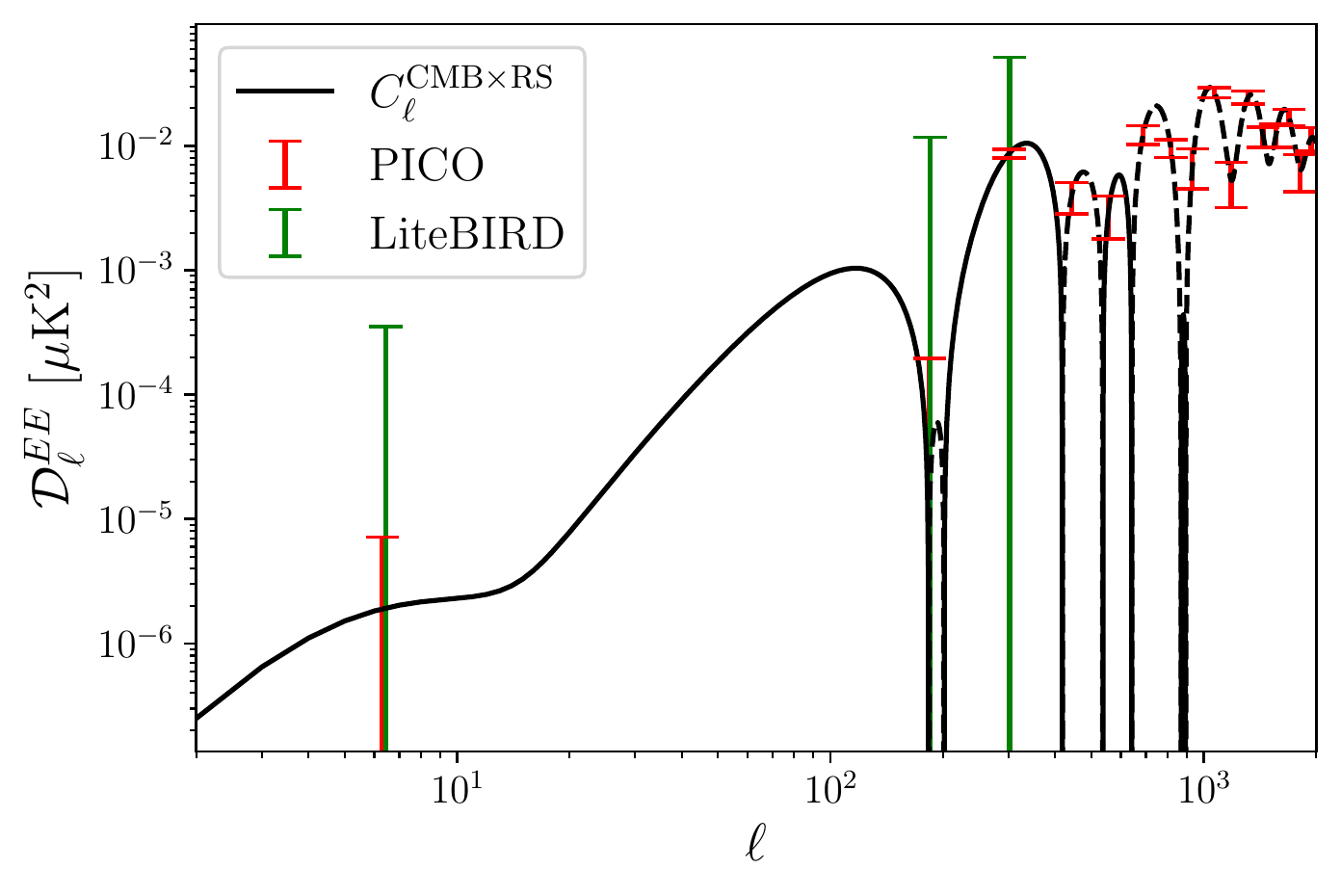}}
  
\caption{Noise comparison for PICO and LiteBIRD. PICO has much tighter error bars, especially on small scales, which will enable a high significance measurement of Rayleigh scattering. LiteBIRD has larger error bars, but can still contribute information on large scales.
}
\label{sp}
\end{figure*}

The results are summarized in Table.~\ref{table}. The small aperture of LiteBIRD restricts the range of scales it can probe to $\ell\lesssim 400$. Including realistic foregrounds, LiteBIRD could reach a $1.9\sigma$ detection in $TT$, which is better than the combined configuration of CCAT-prime, SO, and \textit{Planck}, but insufficient for a significant first detection. PICO would lead to a significant detection even in the presence of foregrounds. With its larger aperture, low noise, and multiple frequency channels that span the millimeter to sub-millimeter wavelength range, we forecast that PICO should achieve a $85\sigma$ detection in $TT$ after foregrounds removal. In Fig.~\ref{sp}, we show the error bars for PICO and LiteBIRD. LiteBIRD has much larger error bars than PICO as expected.

\begin{figure}[htp]

  \centering

  \includegraphics[width=\columnwidth]{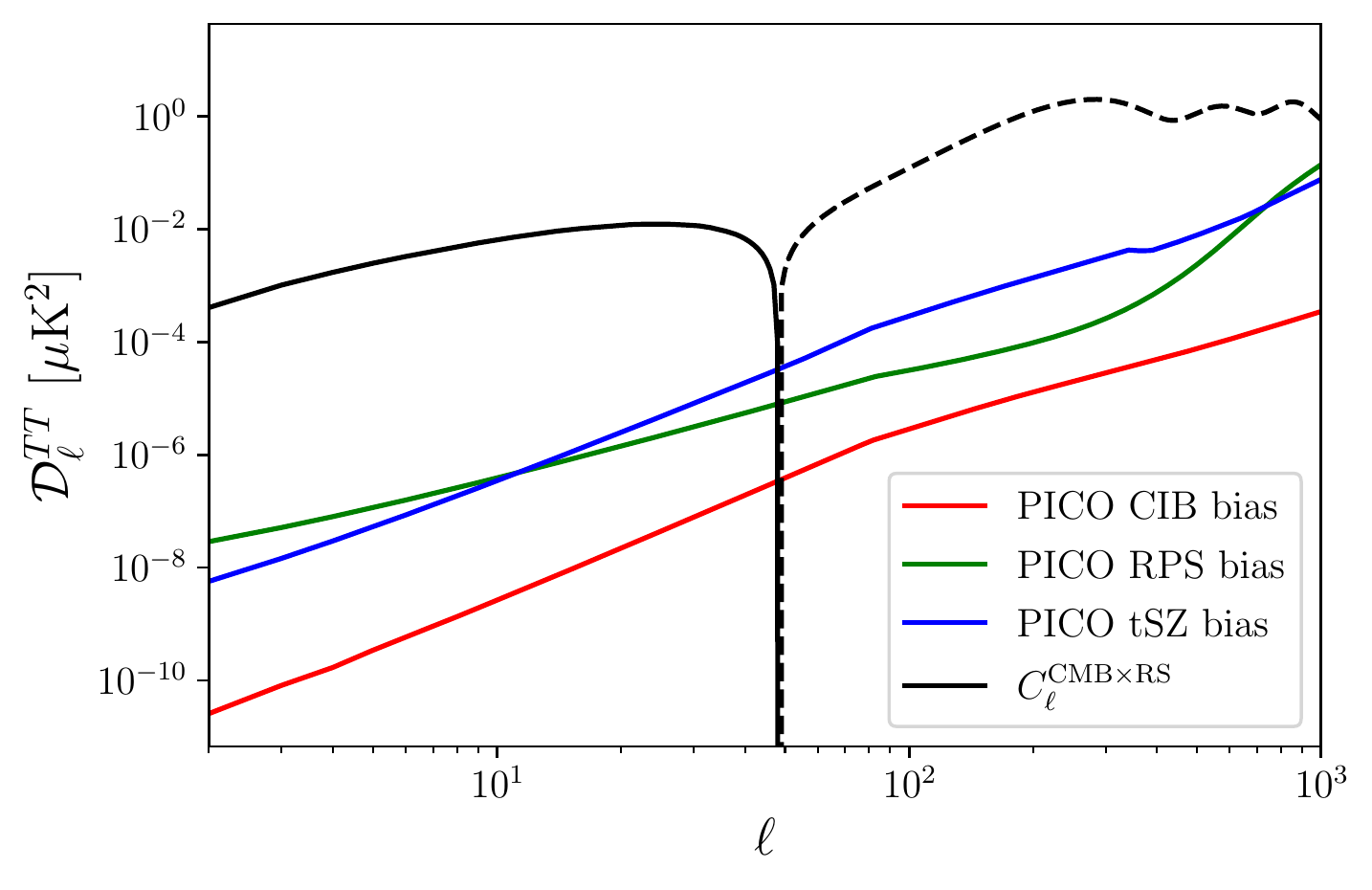}
 
  \caption{Bias of tSZ, CIB, and radio point sources for PICO. The bias of tSZ, CIB and radio point sources are all lower than the CMB-Rayleigh cross-spectrum at $280~\mathrm{GHz}$ by several orders of magnitude.  }
  \label{picobias}
\end{figure}

The fact that PICO is forecast to significantly detect the Rayleigh scattering signal after removing foregrounds, naturally called for a careful study of the biases induced by foregrounds residuals. Following the method outlined in  Sec.~\ref{sec:bias} we calculate the residual foreground bias for PICO. The biases from tSZ, CIB and radio point sources for PICO are shown in Fig.~\ref{picobias}. We find that the CMB-Rayleigh cross-spectrum at $280~\mathrm{GHz}$ is $5$ to $6$ orders of magnitude larger than the bias of tSZ and radio point sources. It is also $8$ to $10$ orders of magnitude larger than the bias of the CIB. It is clear that the multitude of frequency channels and low noise levels not only aid in PICO's ability to detect Rayleigh scattering, but also in its ability to evade residual foreground bias.

\begin{figure}[htp]

  \centering

  \includegraphics[width=\linewidth]{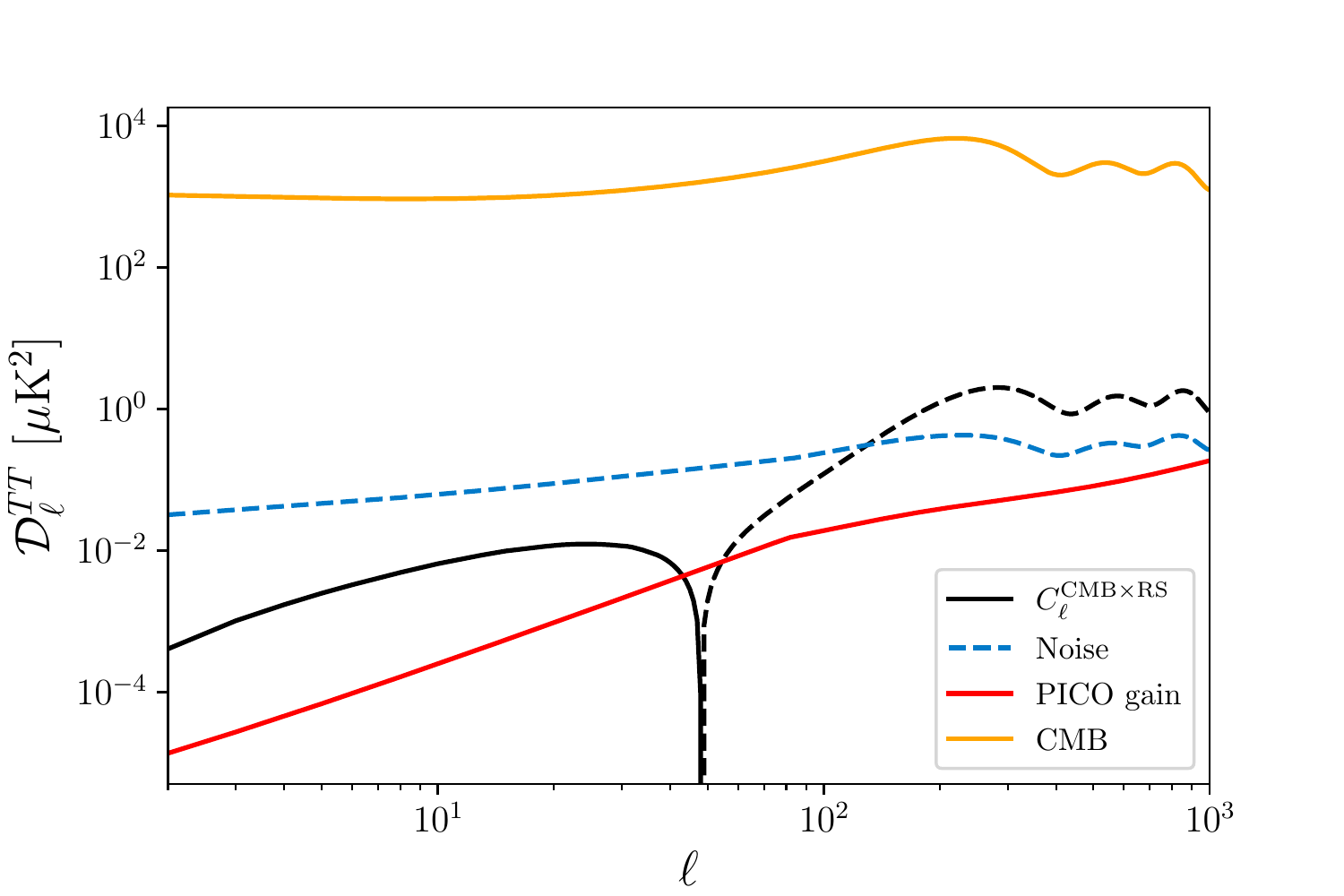}
 
  \caption{Gain of PICO. The gain of $\Delta g=1\%$ of sky emission is around $3$ to $4$ orders of magnitude lower than the noise power spectrum. This means that a $1\%$ gain calibration would be sufficient for PICO.
  }
  \label{picogain}
\end{figure}

The highly significant detection of Rayleigh scattering forecast for PICO additionally leads to the question of calibration uncertainty. Following Sec.~\ref{sec:gain} we calculate the gain bias for a $\Delta g=1\%$. The results are shown in Fig.~\ref{picogain}. At the 1\% threshold the gain bias for PICO is $1$ to $4$ orders of magnitude lower than the reconstructed noise at $\ell < 3 <3000$. This means that a 1\% overall calibration is more than sufficient to safely detect Rayleigh scattering without significant biases. Overall, PICO is robust against the systematics studied here for its forecast detection of the Rayleigh scattering signal.

Given LiteBIRD's low forecast SNR we do not calculate the residual foreground and gain biases for it. However, LiteBIRD's sensitivity on large scale could still provide important information when combined with other measurements. We consider the combination LiteBird and \textit{Planck} with CCAT-prime. 

LiteBIRD contributes no information for $\ell\gtrsim 400$. On large scales $\ell<400$, the reconstructed noise converges to LiteBIRD noise. If we limit the sky fraction to the patch observed by CCAT-prime ($f_\mathrm{sky}=0.4$), we will lose essential information on large scales from LiteBIRD ($f_\mathrm{sky}=0.7$). We calculate the signal-to-noise gained by each telescope in different sky fractions and add them together in quadrature. Essentially, we assign $0$ weight to CCAT-prime in $\ell\leq400$, and $0$ weight to LiteBIRD in $\ell>400$, which is consistent with the cILC method, i.e., 
\begin{align}
     (\mathrm{SNR})^2 & =  \sum_{\ell=2}^{400}\left(\frac{\left(C_{\ell}^{\mathrm{CMB} \times \mathrm{RS}}\right)^2}{\Delta C_{\ell}^2}\right)_\mathrm{LiteBIRD} \nonumber \\
    & + \sum_{\ell=401}^{3000}\left(\frac{\left(C_{\ell}^{\mathrm{CMB} \times \mathrm{RS}}\right)^2}{\Delta C_{\ell}^2}\right)_\mathrm{CCATp+\textit{Planck}}
\end{align}

With the addition of CCAT-prime and \textit{Planck}, the SNR is $2.2$ for the $TT$ signal in the presence of foregrounds, which is $12.5\%$ better than the individual LiteBIRD configuration.

\section{Conclusion}

The observation of CMB Rayleigh scattering has the potential to improve constraints on various cosmological parameters, such as the sum of neutrino masses ($\sum m_{\nu}$), the primordial helium fraction ($Y_\mathrm{He}$), and the light relic density ($\Neff$)~\cite{Alipour2014,Beringue2020}. However, before one can utilize the full potential of the Rayleigh scattering signal for cosmological parameter inference, a first detection is needed. In this paper, we forecast the detectability of the Rayleigh signal in the presence of foregrounds, considering CCAT-prime, Simons Observatory, \textit{Planck}, PICO and LiteBIRD, and applying a cILC to mitigate various foregrounds contamination. We explored combinations of those experiments to minimize the reconstructed noise and determine the effectiveness of these configurations. 

Our analysis shows that in the presence of foregrounds a combination of CCAT-prime, SO and \textit{Planck} can reach an SNR of $1.1$ for the $TT$ CMB-Rayleigh cross-spectrum. As expected, this is below previous estimates which were derived in the absence of foregrounds. This is a clear demonstration that foreground removal, as well as atmospheric contamination, have a large impact on detectability. We compared the capability of these ground-based experiments to future satellite missions. PICO is capable of an $85\sigma$ detection of the $TT$ CMB-Rayleigh cross-spectrum even in the presence of foregrounds. LiteBIRD can potentially achieve an SNR of $1.9$ of the same cross signal. When combined with \textit{Planck} and CCAT-prime, LiteBIRD's SNR is improved to $2.2$. We found that the signal-to-noise of $EE$ cross-spectrum is typically the second highest among all four cross-correlations. While the $EE$  spectrum has the weakest signal strength, it also has the lowest cILC reconstruction noise, since polarized foregrounds are weaker. This also explains why the $EE$ signal-to-noise is least impacted by the presence of foregrounds.

Looking beyond the SNR of Rayleigh scattering, we calculated the biases introduced by tSZ, CIB, and radio point source residuals for the combined configuration of CCAT-prime, SO, and \textit{Planck} and for PICO. For CCAT-prime, SO, and \textit{Planck} the biases from tSZ and CIB have a similar magnitude or are larger than the Rayleigh signal at $\nu=280~\mathrm{GHz}$. Thus a detection of Rayleigh scattering requires further foreground cleaning, for example the deprojection of additional foreground components. Similarly, a denser and broader frequency coverage would help. Indeed for PICO, we found that none of those foregrounds introduce a significant bias which would impact a detection.

Furthermore, we estimate the required instrumental calibration to prevent errors in the gain that may impact the detectability of Rayleigh scattering. We found that an overall gain calibration error of $\Delta g=1\%$ is sufficient for CCAT-prime, SO, and \textit{Planck} and  PICO. Nevertheless, a more comprehensive study of the calibration of CCAT-prime, SO, and PICO is still warranted and practical for upcoming analysis.

With CCAT-prime, SO, and \textit{Planck}, a first detection is unlikely to be achievable by relying solely on constrained-ILC methods. Due to the blind nature of such methods, they do not allow us to leverage the scale dependence of the Rayleigh scattering signal. Indeed, for a fixed cosmology, the Rayleigh scattering signal is straightforward to model, which would be an incentive to explore more parametric approaches such as maximum likelihood algorithm for parametric component separation \citep{stompor_maximum_2009, abylkairov_partially_2021}. More realistic analysis of instrumental calibration requirements will also be necessary and will be the topic of future investigations. Telescopes not considered here, such as SPT \citep[][]{SPT3G}, CMB-S4 \citep[][]{CMBS4} and CMB-HD \citep[][]{CMBHD}, are guaranteed to help in an attempt to make a first detection of CMB Rayleigh scattering. 

The detection of the Rayleigh signal is an exciting prospect which will eventually allow us to observe a frequency-dependent last-scattering surface. With advancing telescope technology and new data analysis techniques, there is hope to make a first detection of Rayleigh signal this decade. Eventually a significant detection will allow us to use the cosmological information hidden in the Rayleigh signal to improve parameter constraints beyond those achievable from the primary CMB alone.

\section*{Acknowledgements}
We thank Jens Chluba, Tom Crawford, and Blake Sherwin for their useful comments on our manuscript. NB acknowledges support from NSF grant AST-1910021 and NASA grants 21-ADAP21-0114 and 21-ATP21-0129. PDM acknowledges support from the Netherlands organisation for scientific research (NWO) VIDI grant (dossier 639.042.730). JM is supported by the US~Department of Energy under Grant~\mbox{DE-SC0010129}. BB acknowledges postdoctoral support from the European Research Council (ERC) under the European Union’s Horizon 2020 research and innovation programme (Grant agreement No. 849169). SKC acknowledges support from NSF award AST-2001866.

\appendix

\section{Experimental configurations}
\label{sec:Telescopes}

The following tables summarize the noise models for all experiments considered in this work. More details can be found in Sec.~\ref{sec:Tele}.

\begin{table}[H]
\centering
SO LAT ($f_\mathrm{sky}=0.4$,  $\ell_\mathrm{knee}^T=1000$, $\alpha_\mathrm{knee}^T=-3.5$, $\ell_\mathrm{knee}^P=700$, $\alpha_\mathrm{knee}^P=-1.4$)
\begin{tabular}{|c|c|c|c|}

\hline
\rowcolor[HTML]{EBFFDF} Freq (GHz) & Beam (arcmin) & $N_\mathrm{white}$($\mu \mathrm{K}^2$) & $N_\mathrm{red}$($\mu \mathrm{K}^2s$) \\ \hline
\cellcolor[HTML]{EBFFDF} 27 & 7.4 & $2.3\times10^{-4}$  & 100  \\ \hline
\rowcolor[HTML]{EFEFEF} \cellcolor[HTML]{EBFFDF}39 & 5.1  & $6.2\times10^{-5}$  & 39  \\ \hline
\cellcolor[HTML]{EBFFDF}93 & 2.2 & $2.8\times10^{-6}$  & 230  \\ \hline
\rowcolor[HTML]{EFEFEF} \cellcolor[HTML]{EBFFDF}145 & 1.4  & $3.6\times10^{-6}$  & 17000  \\ \hline
\cellcolor[HTML]{EBFFDF}225 & 1.0  & $1.9\times10^{-5} $ & 17000  \\ \hline
\rowcolor[HTML]{EFEFEF} \cellcolor[HTML]{EBFFDF}280 & 0.9  & $1.16\times10^{-4}$  & 31000  \\ \hline

\end{tabular}
\caption{Survey configuration of Simons Observatory Large Aperture Telescope (SO LAT) taken from \citet[][]{SOforecasts}.}
\label{tab:SO_LAT}
\end{table}

\begin{table}[H]
\centering
SO SAT ($f_\mathrm{sky}=0.1$)

\begin{tabular}{|c|c|c|c|c|}

\hline
\rowcolor[HTML]{EBFFDF} Freq (GHz) & Beam (arcmin) & $N_\mathrm{white}=N_\mathrm{red}$($\mu \mathrm{K}^2$) & $\ell_\mathrm{knee}^P$ &$\alpha_\mathrm{knee}^P$ \\ \hline
\cellcolor[HTML]{EBFFDF} 27 & 91 & $5.3\times10^{-5}$  & 15 &$-2.4$  \\ \hline
\rowcolor[HTML]{EFEFEF} \cellcolor[HTML]{EBFFDF} 39 & 63  & $2.4\times10^{-5}$  &15 & $-2.4$ \\ \hline
\cellcolor[HTML]{EBFFDF} 93 & 30 & $3.0\times10^{-7}$  & 25 &$-2.5$ \\ \hline
\rowcolor[HTML]{EFEFEF} \cellcolor[HTML]{EBFFDF} 145 & 17  & $3.7\times10^{-7}$  & 25 &$-3.0$ \\ \hline
\cellcolor[HTML]{EBFFDF} 225 & 11  & $1.5\times10^{-6} $ & 35  &$-3.0$ \\ \hline
\rowcolor[HTML]{EFEFEF} \cellcolor[HTML]{EBFFDF} 280 & 9  & $8.5\times10^{-6}$  & 40  &$-3.0$ \\ \hline

\end{tabular}

\caption{Survey configuration of Simons Observatory Small Aperture Telescope (SO SAT) taken from \citet[][]{SOforecasts}.}
\label{tab:SO_SAT}
\end{table}

\begin{table}[H]
\centering
CCAT-prime ($f_\mathrm{sky} = 0.4$, $\ell_\mathrm{knee}^T=1000$, $\alpha_\mathrm{knee}^T=-3.5$, $\ell_\mathrm{knee}^P=700$, $\alpha_\mathrm{knee}^P=-1.4$)
\begin{tabular}{|c|c|c|c|}

\hline
\rowcolor[HTML]{EBFFDF} Freq (GHz) & Beam (arcsec) & $N_\mathrm{white}$($\mu \mathrm{K}^2$) & $N_\mathrm{red}$($\mu \mathrm{K}^2$) \\ \hline
\cellcolor[HTML]{EBFFDF} 220 & 57 & $1.8\times10^{-5}$  & $1.6\times10^{-2}$  \\ \hline
\rowcolor[HTML]{EFEFEF} \cellcolor[HTML]{EBFFDF} 280 & 45  & $6.4\times10^{-5}$  & $1.1\times10^{-1}$  \\ \hline
\cellcolor[HTML]{EBFFDF} 350 & 35 & $9.3\times10^{-4}$  & $2.7\times10^{0}$  \\ \hline
\rowcolor[HTML]{EFEFEF} \cellcolor[HTML]{EBFFDF} 410 & 30  & $1.2\times10^{-2}$  & $1.7\times10^{1}$  \\ \hline
\cellcolor[HTML]{EBFFDF} 850 & 14 & $2.8\times10^4 $ & $6.1\times10^{6}$  \\ \hline

\end{tabular}
\caption{Survey configuration of CCAT-prime taken from \citet{CCATp}.}
\label{tab:CCATp}
\end{table}

\vspace*{0.2cm}

\begin{table}[H]
\centering
Planck ($f_\mathrm{sky}=0.6$)

\begin{tabular}{|c|c|c|}

\hline
\rowcolor[HTML]{EBFFDF} Freq (GHz) & Beam (arcmin) & $N_\mathrm{white}$($\mu \mathrm{K}^2$) \\ \hline
\cellcolor[HTML]{EBFFDF} 100 & 9.7 & $5.07\times10^{-4}$ \\ \hline
\rowcolor[HTML]{EFEFEF} \cellcolor[HTML]{EBFFDF} 143 & 7.2 & $9.21\times10^{-5}$ \\ \hline

\cellcolor[HTML]{EBFFDF} 217 & 4.9 & $1.85\times10^{-4}$ \\ \hline
\rowcolor[HTML]{EFEFEF} \cellcolor[HTML]{EBFFDF} 353 & 4.9 & $2.00\times10^{-3}$
\\ \hline
\end{tabular}

\caption{Survey configuration of Planck taken from \citet[][]{Planck2018}.}
\label{tab:Planck}
\end{table}

\begin{table}[H]
\centering
LiteBIRD ($f_\mathrm{sky}=0.7$)

\begin{tabular}{|c|c|c|}

\hline
\rowcolor[HTML]{EBFFDF} Freq (GHz) & Beam (arcmin) & $N_\mathrm{white}$($\mu \mathrm{K}^2$) \\ \hline
\cellcolor[HTML]{EBFFDF} 40 & 69 & $2.38\times10^{-4}$ \\ \hline
\rowcolor[HTML]{EFEFEF} \cellcolor[HTML]{EBFFDF} 50 & 56 & $9.75\times10^{-5}$ \\ \hline
\cellcolor[HTML]{EBFFDF} 60 & 48 & $6.70\times10^{-5}$ \\ \hline
\rowcolor[HTML]{EFEFEF} \cellcolor[HTML]{EBFFDF} 68 & 43 & $4.44\times10^{-4}$ \\ \hline
\cellcolor[HTML]{EBFFDF} 78 & 39 & $3.08\times10^{-5}$ \\ \hline
\rowcolor[HTML]{EFEFEF} \cellcolor[HTML]{EBFFDF} 89 & 35 & $2.32\times10^{-5}$ \\ \hline
\cellcolor[HTML]{EBFFDF} 100 & 29 & $1.43\times10^{-5}$ \\ \hline
\rowcolor[HTML]{EFEFEF} \cellcolor[HTML]{EBFFDF} 119 & 25 & $9.77\times10^{-6}$ \\ \hline
\cellcolor[HTML]{EBFFDF} 140 & 23 & $5.89\times10^{-6}$ \\ \hline
\rowcolor[HTML]{EFEFEF} \cellcolor[HTML]{EBFFDF} 166 & 21 & $7.15\times10^{-6}$ \\ \hline
\cellcolor[HTML]{EBFFDF} 195 & 40 & $5.69\times10^{-6}$ \\ \hline
\rowcolor[HTML]{EFEFEF} \cellcolor[HTML]{EBFFDF} 235 & 19 & $1.00\times10^{-5}$ \\ \hline
\cellcolor[HTML]{EBFFDF} 280 & 24 & $2.95\times10^{-5}$ \\ \hline
\rowcolor[HTML]{EFEFEF} \cellcolor[HTML]{EBFFDF} 337 & 20 & $6.44\times10^{-5}$ \\ \hline
\cellcolor[HTML]{EBFFDF} 402 & 17 & $2.38\times10^{-4}$ \\ \hline
\end{tabular}

\caption{Survey configuration of LiteBIRD taken from \citet[][]{LiteBird}.}
\label{tab:LiteBIRD}
\end{table}

\begin{table}[H]
\centering
  PICO ($f_\mathrm{sky}=0.7$)

\begin{tabular}{|c|c|c|}

\hline
\rowcolor[HTML]{EBFFDF} Freq (GHz) & Beam (arcmin) & $N_\mathrm{white}$($\mu \mathrm{K}^2$) \\ \hline
\cellcolor[HTML]{EBFFDF} 21 & 38.4 & $1.21\times10^{-5}$ \\ \hline
\rowcolor[HTML]{EFEFEF} \cellcolor[HTML]{EBFFDF} 25 & 32.0 & $7.15\times10^{-6}$ \\ \hline
\cellcolor[HTML]{EBFFDF} 30 & 28.3 & $3.20\times10^{-6}$ \\ \hline
\rowcolor[HTML]{EFEFEF} \cellcolor[HTML]{EBFFDF} 36 & 23.6 & $1.33\times10^{-6}$ \\ \hline
\cellcolor[HTML]{EBFFDF} 43 & 22.2 & $1.33\times10^{-6}$ \\ \hline
\rowcolor[HTML]{EFEFEF} \cellcolor[HTML]{EBFFDF} 52 & 18.4 & $6.77\times10^{-7}$ \\ \hline
\cellcolor[HTML]{EBFFDF} 62 & 12.8 & $6.11\times10^{-7}$ \\ \hline
\rowcolor[HTML]{EFEFEF} \cellcolor[HTML]{EBFFDF} 75 & 10.7 & $3.81\times10^{-7}$ \\ \hline
\cellcolor[HTML]{EBFFDF} 90 & 9.5 & $1.69\times10^{-7}$ \\ \hline
\rowcolor[HTML]{EFEFEF} \cellcolor[HTML]{EBFFDF} 108 & 7.9 & $1.08\times10^{-7}$ \\ \hline
\cellcolor[HTML]{EBFFDF} 129 & 7.4 & $9.52\times10^{-8}$ \\ \hline
\rowcolor[HTML]{EFEFEF} \cellcolor[HTML]{EBFFDF} 159 & 6.2 & $7.15\times10^{-8}$ \\ \hline
\cellcolor[HTML]{EBFFDF} 186 & 4.3 & $3.32\times10^{-7}$ \\ \hline
\rowcolor[HTML]{EFEFEF} \cellcolor[HTML]{EBFFDF} 223 & 3.6 & $4.33\times10^{-7}$ \\ \hline
\cellcolor[HTML]{EBFFDF} 268 & 3.2 & $2.05\times10^{-7}$ \\ \hline
\rowcolor[HTML]{EFEFEF} \cellcolor[HTML]{EBFFDF} 321 & 2.6 & $3.81\times10^{-7}$ \\ \hline
\cellcolor[HTML]{EBFFDF} 385 & 2.5 & $4.33\times10^{-7}$ \\ \hline
\rowcolor[HTML]{EFEFEF} \cellcolor[HTML]{EBFFDF} 462 & 2.1 & $1.73\times10^{-6}$ \\ \hline
\cellcolor[HTML]{EBFFDF} 555 & 1.5 & $4.44\times10^{-5}$ \\ \hline
\rowcolor[HTML]{EFEFEF} \cellcolor[HTML]{EBFFDF} 666 & 1.3 & $6.61\times10^{-4}$ \\ \hline
\cellcolor[HTML]{EBFFDF} 799 & 1.1 & $2.32\times10^{-2}$ \\ \hline
\end{tabular}
\caption{Survey configuration of PICO taken from \citet[][]{PICO}.}
\label{tab:PICO}
\end{table}

\bibliographystyle{apj}

\bibliography{yz}

\end{document}